\documentclass{amsart}

\usepackage{epsfig}
\usepackage{latexsym}
\usepackage{amssymb}
\usepackage{amsmath}
\usepackage{euler}

\def\cutmeasure{\gamma}
\def\dimlev{d}
\def\dedrule#1#2#3{#1 \ \frac{\displaystyle #2}{\displaystyle #3}}
\def\ie{\emph{i.e.}}
\def\linimpl{\multimap}
\def\linear{\multimap}
\def\llpar{\wp}
\def\lblmeasure{\mu}
\def\maximaldimension{\boldsymbol{D}}
\def\maximaldepth{\boldsymbol{\partial}}
\def\mmm{\maximaldepth}
\def\maximalweight{\boldsymbol{W}}
\def\rewrel{\triangleright}
\def\rewl{\rewrel_l}
\def\rews{\rewrel_s}
\def\rewp{\rewrel_p}
\def\sug#1#2{\{{}^{#1}\!\downarrow{}_{#2}\}}
\def\strategy{\rewrel_{\sigma}}
\def\weight{\operatorname{wgt}}

\newtheorem{definition}{Definition}

\newtheorem{fact}{Fact}

\newtheorem{proposition}{Proposition}
\newenvironment{nonumtheorem}
{\vspace{\baselineskip}\textbf{Theorem.}}
{\vspace{\baselineskip}}
\newenvironment{remark}
{\begin{trivlist}\item[]\textbf{Remark.}}{\end{trivlist}}
\newtheorem{theorem}{Theorem}

\newcommand{\A}{\alpha}
\newcommand{\B}{\beta}
\newcommand{\C}{\gamma}

\newcommand{\DB}{\overline{!}}
\newcommand{\DP}{\Bar{\phantom{|}\S}}
\newcommand{\FV}[1]{{\mathtt{FV}} (#1)}
\newcommand{\msug}[4]
{\{^{#1}\!\!\downarrow\! _{#2}\cdots ^{#3}\!\!\downarrow\! _{#4}\}}
\newcommand{\proves}{\vdash}
\newcommand{\rewsyst}{\leadsto}
\newcommand{\tensor}{\!\otimes\!}
\newcommand{\Terms}{\Lambda}
\newcommand{\TermVars}{\mathtt{T_{variables}}}

\newcommand{\TS}{\proves}
\newcommand{\VS}{\alpha}

\newcommand{\Base}{\mathit{base}}
\newcommand{\Coerc}{\mathit{coerc}}
\newcommand{\doublelangle}{\langle\!\langle}
\newcommand{\doublerangle}{\rangle\!\rangle}
\newcommand{\Fst}{\mathit{fst}}
\newcommand{\Int}{\boldsymbol{Int}}
\newcommand{\IntT}{\boldsymbol{Int}}
\newcommand{\Iter}{\mathit{iter}}
\newcommand{\Mult}{\mathit{mult}}
\newcommand{\Num}[1]{\overline{#1}}
\newcommand{\Pred}{\mathit{pred}}
\newcommand{\Snd}{\mathit{snd}}
\newcommand{\Step}{\mathit{step}}
\newcommand{\Succ}{\mathit{succ}}
\newcommand{\Sum}{\mathit{sum}}
\newcommand{\Tuple}{\mathit{tuple}}
\newcommand{\Zero}{\overline{\mathit{0}}}

\newcommand{\ABool}{\mathbf{proj}}
\newcommand{\BaseT}{\mathbf{base}}

\newcommand{\Comp}{\mathit{next\_config}}
\newcommand{\CompT}{\mathbf{next\_config}}
\newcommand{\Conf}{\mathit{config}}
\newcommand{\ConfT}{\mathbf{config}}
\newcommand{\Conftoconf}{\mathit{config2config}}
\newcommand{\Conftotape}{\mathit{config2tape}}
\newcommand{\Coercinitconf}{\mathit{coerc\_init\_config}}
\newcommand{\Doubletape}{\mathit{dbl\_tape}}
\newcommand{\Emptytape}{\mathit{empty\_tape}}
\newcommand{\Emptyinitconf}{\mathit{empty\_init\_config}}
\newcommand{\Ib}{\mathbf{\multimap}}
\newcommand{\Iterint}{\mathit{iter}}
\newcommand{\Left}{\mathit{left}}
\newcommand{\Pib}{\Pi_\bot}
\newcommand{\Pie}{\Pi_\emptyset}
\newcommand{\Pio}{\Pi_1}
\newcommand{\Pis}{\Pi_\star}
\newcommand{\Pit}{\Pi_\top}
\newcommand{\Piz}{\Pi_0}
\newcommand{\Right}{\mathit{right}}
\newcommand{\RowT}{\mathbf{row}}
\newcommand{\ShiftT}{\mathbf{shift}}
\newcommand{\Stay}{\mathit{stay}}
\newcommand{\State}{\mathit{state}}
\newcommand{\StateT}{\mathbf{state}}
\newcommand{\StepT}{\mathbf{step}}
\newcommand{\Succtape}{\mathit{succ\_tape}}
\newcommand{\Succinitconf}{\mathit{succ\_init\_config}}

\newcommand{\sfs}{\mathsf{s}}
\newcommand{\TapeT}{\mathbf{tape}}
\newcommand{\Tfun}{\trns{\transitionf}}
\newcommand{\TfunT}{\boldsymbol{\trns{\transitionf}}}
\newcommand{\transitionf}{\delta}
\newcommand{\trns}[1]{\widehat{#1}}
\newcommand{\Tapetoint}{\mathit{tape2int}}
\newcommand{\Tapetoconf}{\mathit{tape2config}}
\newcommand{\Tapetoinitconf}{\mathit{tape2init\_config}}

\begin{document}
\title[Light Affine Logic]
{Light Affine Logic\\
(\tiny Proof Nets,
Programming Notation,\\
P-Time Correctness and Completeness)}
\author{Andrea Asperti and Luca Roversi}
\address{
Andrea Asperti\\
Dipartimento di Scienze dell'Informazione\\
Via di Mura Anteo Zamboni, n. 7\\
40127  Bologna -- ITALY\\
\textsc{e-mail}: asperti@cs.unibo.it
}
\address{
Luca Roversi\\ 
Dipartimento di Informatica\\
C.so Svizzera, n. 185\\
10149 Torino -- ITALY\\
\textsc{e-mail}: rover@di.unito.it
}
\date{\today}

\begin{abstract}
This paper is a structured introduction to Light Affine Logic,
and to its intuitionistic fragment. Light Affine Logic has a polynomially
costing cut elimination (P-Time correctness), and encodes all P-Time Turing
machines (P-Time completeness).
P-Time correctness is proved by introducing the Proof nets for Intuitionistic
Light Affine Logic.
P-Time completeness is demonstrated in full details thanks to a
very compact program notation. On one side, the proof of P-Time
correctness describes how the complexity of cut elimination is
controlled, thanks to a suitable cut elimination strategy that exploits
structural properties of the Proof nets.
This allows to have a good catch on the meaning of the $\S$ modality, which is
a peculiarity of light logics.
On the other side, the proof of P-Time completeness, together with a lot of 
programming examples,
gives a flavor of the non trivial task of programming with resource
limitations, using Intuitionistic Light Affine Logic derivations as programs.
\end{abstract}

\maketitle

\section{Introduction}

This paper belongs to the area of
polytime computational systems \cite{GSS92,LM93,Le94,Girard:LLL98}.
The purpose of such systems is manifold. On the theoretical side, 
they provide a better understanding about the 
\emph{logical essence} of calculating with time restrictions.
On the practical side, via the Curry-Howard correspondence
\cite{GLT:PT}, they yield sophisticated
typing systems that, \emph{statically}, provide  an 
accurate upper bound on the complexity of the computation.
The types give essential information
on the strategy  to efficiently reduce the terms they type.

A cornerstone in the area is 
Girard's Light Linear Logic \cite{Girard:LLL98} (LLL),
a deductive system with cut elimination, \ie\ a logical system.
In \cite{Asperti:LICS98}, Light Affine Logic (LAL), a slight variation of LLL,
was introduced. In \cite{Roversi:CSL99} there are some basic observations
about how P-Time completeness of LAL, and, in fact, of LLL as well, can be
proved. This paper is a monolithic reworking of both papers
with the hope to make the subject more widely accessible. It must be clear,
however, that the paper is addressed to people already acquainted with the
basic notions of Linear Logic \cite{Gi95}.

The main results of this paper are two theorems about Intuitionistic Light
Affine Logic (ILAL).

\begin{nonumtheorem}
Every derivation $\Pi$ of ILAL can be transformed
into its cut free form in a number of cut elimination steps bound by a
polynomial in the dimension of $\Pi$.
\end{nonumtheorem}

We shall see that the degree of the polynomial is an exponential function of
the \emph{depth} of $\Pi$. The meaning of ``depth'' will become clearer later,
but we can already say that it is a purely proof-theoretic structural notion.

\begin{nonumtheorem}
Every P-Time Turing machine can be encoded and simulated by a derivation of
ILAL.
\end{nonumtheorem}

The two theorems together imply that ILAL is a logical system,
equivalent to the set of P-Time Turing machines, with respect to the
cost and to the expressivity.

In more details, LAL is introduced by adding full weakening to LLL.
This modification, while not altering the good complexity
property, greatly simplifies the logical system.
Firstly, the number of rules decreases
from 21 to just 11 rules, with respect to LLL.
Secondly, LAL is endowed with additives, without adding them explicitly:
in presence of weakening, their computational behavior is there for free.
This point will become clear later, when encoding the predecessor on Church
numerals, and some components of P-Time Turing machines.

Rephrasing Girard \cite{Girard:LLL98}, the slogan behind the design of LAL is:
\emph{the abuse of contraction may have damaging complexity
effects, but the abstinence from weakening leads to
inessential syntactical complications}.

	\subsection{Light Affine Logic}

As we said, LAL is both a variant, and a simplification of LLL.
The main intuitions about the new modalities
of LLL are preserved by their counterparts
of LAL. We recall them here below.
Let ${\mathcal T}$ be the set of \emph{literals} in
Figure~\ref{figure:LAL-literals}. 
\begin{figure}[htbp]
\begin{eqnarray*}
{\mathcal T} =\{\alpha      , \beta      , \gamma      ,\dots, 
               \alpha^\perp, \beta^\perp, \gamma^\perp,\dots \}
\end{eqnarray*}
\caption{Literals of LAL}
\label{figure:LAL-literals}
\end{figure}

The set $\mathcal F$ of formulas, is defined in two steps.
Firstly, consider the language generated by the grammar in
Figure~\ref{figure:LAL-formulas}.
\begin{figure}[htbp]
\[
\begin{array}{rcl}
A &::=& {\mathcal T}\ \mid\ A \otimes A\ \mid\ A \llpar A\\
  &   & \forall \alpha. A\ \mid\ \exists \alpha. A\\
  &   & !A\ \mid\ ?A \ \mid\ \S A
\end{array}
\]
\caption{Formulas of LAL}
\label{figure:LAL-formulas}
\end{figure}
Secondly, partition such a language into equivalence classes
by the negation $(\ )^\perp$, defined in
Figure~\ref{figure:LAL-negation}.
\begin{figure}[htbp]
\begin{eqnarray*}
\begin{array}{rcl}
(\alpha)^\perp       &=& \alpha^\perp\\
(\alpha^\perp)^\perp &=& \alpha\\
(!A)^\perp           &=& ?(A^\perp)\\
(\S A)^\perp         &=& \S (A^\perp)\\
(?A)^\perp           &=& !(A^\perp)\\
(A \otimes B)^\perp  &=& A^\perp \llpar B^\perp\\
(A \llpar B)^\perp   &=& A^\perp \otimes B^\perp\\
(\forall \alpha.A)^\perp &=& \exists \alpha. A^\perp\\
(\exists \alpha.A)^\perp &=& \forall \alpha. A^\perp
\end{array}
\end{eqnarray*}
\caption{Negation on the formulas}
\label{figure:LAL-negation}
\end{figure}

The sequent calculus of (classical) LAL is in Figure~\ref{figure:LAL}.
\begin{figure}[htbp]
\begin{tabular}{ll}
$\dedrule{(Ax)}
{}
{\vdash A,A^{\bot}}$
&
$\dedrule{(Cut)}
{\vdash \Gamma,A 
 \quad
 \vdash A^\bot,\Delta}
{\vdash \Gamma, \Delta}$
\\ \\
$\dedrule{(Perm.)}
{\vdash \Gamma,A,B,\Delta}
{\vdash \Gamma,B,A,\Delta}$
\\ \\ 
$\dedrule{(Contr.)}
{\vdash \Gamma,?A,?A}
{\vdash \Gamma,?A} $
&
$\dedrule{(Weak.)}
{\vdash \Gamma}
{\vdash \Gamma,A}$
\\ \\
$\dedrule{(\otimes)}
{\vdash \Gamma,A \hspace{1cm}\vdash B,\Delta}
{\vdash \Gamma,A\otimes B,\Delta}$
& 
$\dedrule{(\llpar)}
{\vdash \Gamma,A,B}
{\vdash \Gamma, A\llpar B}$
\\ \\
$\dedrule{(!)}
{\vdash B,A}
{\vdash ?B,!A} $
&
$\dedrule{(\S)}
{\vdash B_1,\dots, B_n,A_1,\dots, A_m}
{\vdash ?B_1,\dots, ?B_n,\S A_1,\dots, \S A_m}$
\\ \\
$\dedrule{(\forall)}
{\vdash \Gamma,A}
{\vdash \Gamma,\forall . \alpha A}\;\;(\alpha \not\in FV(\Gamma))\hspace{1cm}$
&
$\dedrule{(\exists)}
{\vdash \Gamma,A[^B/_\alpha]}
{\vdash \Gamma,\exists .\alpha A}$
\end{tabular}
\caption{Light Affine Logic}
\label{figure:LAL}
\end{figure}
Observe that $B$ can be absent in rule $(!)$, and that the sequence
$B_1,\dots,B_n$ of rule $(\S)$ can be empty.

Like in Linear
Logic, we may only perform contraction (dually, duplication) on variables 
of type ?A (dually, data of type !A). 
However, in LAL, and in LLL, the potential explosion 
of the computation, essentially due to an explosion of the use
of the rule $(Contr.)$, also called \emph{sharing}
\cite{AG98}, is taken under control.
This is achieved by constraining the
!-boxes to have at most one input (see $(!)$-rule).
So, the number of sharing structures,
\ie, of contraction rules, cannot grow while duplicating a !-box.
This limitation enormously decreases the overall expressivity.
It is recovered by adopting a self-dual modality $\S$, which corresponds to
introducing $\S$-boxes in the derivations of LAL.
A $\S$-box may contain several shared (dually, contracted) variables
(\ie, \emph{multiple} occurrences of ?-assumptions).
However, in this case, the $\S$-box itself cannot be duplicated
to prevent the explosion of sharing.

The \emph{key point} is that, adding unrestricted weakening to LLL,
does not violate these complexity intuitions!

The basic logical problem with LAL is the elimination of the
cut between $\vdash \Gamma,A$ and $\vdash \Delta,A^{\bot}$ when both 
$A$, and $A^{\bot}$ are immediately
introduced by a weakening, which is also the usual problem with
interpretations of cut elimination as computation in classical logic.

We shall simply avoid this problem by restricting our attention to 
the intuitionistic fragment ILAL of LAL.

Section~\ref{section:Intuitionistic Light Affine Logic} recalls the sequent
calculus of ILAL.
Section~\ref{section:Proof Nets} introduces the graph language of Proof
nets for ILAL, with some terminology.
Section~\ref{section:Cut Elimination on Proof Nets} is about the cut elimination
step on Proof nets.
Section~\ref{section:P-Time Correctness} develops the proof of P-Time
correctness. The proof is classical: we supply strictly decreasing measures
as the cut elimination proceeds.
Section~\ref{section:The Concrete Syntax} defines the functional language that
realizes (a sort of) Curry-Howard isomorphism for ILAL. It is the first step
towards the proof of P-Time completeness. 
Section~\ref{section:The Type Assignment} decorates the sequent calculus
derivations of ILAL with the terms of the functional language, so using the
sequent calculus as a type assignment. The relation derivation/term is not
one-to-one. This is why our instance of Curry-Howard isomorphism is not,
in fact, a true isomorphism. This will not constitute any problems, as
discussed in Section~\ref{section:Comments on the Concrete Syntax},
once introduced the dynamics of the functional language in
Section~\ref{section:The Dynamics Concrete Syntax}. Obviously, the dynamics
is, more or less, a restatement of the cut elimination steps in the functional
syntax. Section~\ref{section:Numerical System} is the first programming example
with our functional notation. We develop a numerical system with a predecessor
which is syntactically linear, up to weakening, and which obeys a general
programming scheme, that we will sometimes exploit to encode the whole class of
P-Time Turing machines as well.
This is the second step towards P-Time completeness proof.
Section~\ref{section:Polynomials} contains a second programming example. For the
first time, we write all the details to encode the polynomials with positive
degree and positive coefficients as derivations of ILAL.
Section~\ref{section:P-Time Completeness} proves P-Time completeness.
The proof is a further programming exercise. It consists of the definition of a
translation from P-Time Turing machines to terms of our functional language.
For a simpler encoding, we make some simplifying, but not restricting
assumptions, on the class of P-Time Turing machine effectively encoded.
Section~\ref{section:Conclusions} concludes the paper with some observations and
hypothesis on future work.

\section{Intuitionistic Light Affine Logic}
\label{section:Intuitionistic Light Affine Logic}

Intuitionistic LAL (ILAL) is the logical system based on the connectives
$\linear$, $\otimes$, !, $\S$,
and $\forall$ of LAL, where $A\linear B$ is a notation
for $A^\perp\llpar B$. The sequent calculus for ILAL is
in Figure~\ref{figure:ILAL}.
\begin{figure}[htbp]
\begin{tabular}{ll}
$\dedrule{(Ax)}{}{A\vdash A}$
&
$\dedrule{(Cut)}
{\Gamma \vdash A
 \quad
 A,\Delta \vdash B}
 {\Gamma, \Delta \vdash B}$
\\ \\
$\dedrule{(Perm.)}
{\Gamma,A,B,\Delta \vdash C}
{\Gamma, B, A,\Delta \vdash C}$
\\ \\
$\dedrule{(Weak.)}{\Gamma \vdash C}{\Gamma, A \vdash C}$
&
$\dedrule{(Contr.)}
{\Gamma,!A,!A \vdash B}
{\Gamma,!A \vdash B}$
\\ \\
$\dedrule{(\linear_l)}
{\Gamma \vdash A
 \quad
 B,\Delta \vdash C}
{\Gamma,A\linear B,\Delta \vdash C}$
&
$\dedrule{(\linear_r)}
{\Gamma,A \vdash B}
{\Gamma \vdash A\linear B}$
\\ \\ 
$\dedrule{(\otimes_l)}
{\Gamma, A, B \vdash C}
{\Gamma, A\otimes B \vdash C}$
&
$
\dedrule{(\otimes_r)}
{\Gamma \vdash A
 \quad
 \Delta \vdash B}
{\Gamma, \Delta \vdash A\otimes B}$
\\ \\ 
$\dedrule{(!)}
{B \vdash A}
{!B \vdash !A}$
&
$\dedrule{(\S)}
{\Delta,\Gamma \vdash C}
{!\Delta,\S \Gamma \vdash \S C}$
\\ \\
$\dedrule{(\forall_l)}
{\Gamma,A\sug{B}{\alpha} \vdash C}
{\Gamma,\forall \alpha . A \vdash C}$
&
$\dedrule{(\forall_r)}
{\Gamma \vdash A}
{\Gamma \vdash \forall \alpha. A}
 \quad
 (\alpha \not\in FV(\Gamma))$

\end{tabular}
\caption{Intuitionistic Light Affine Logic}
\label{figure:ILAL}
\end{figure}
Like in Classical LAL (Figure~\ref{figure:LAL}), the assumption $B$ of
rule $(!)$ may be absent, and one, or both, of the sets of assumptions
$\Delta$ and $\Gamma$ of rule $(\S)$ may be empty.

Our goal is twofold.
On one side, we want to prove that the cut elimination of the
system here above is \emph{correct} with respect to the class
P-Time. Namely, we want to prove that, given a derivation $\Pi$, it
can be reduced to its normal form, through cut elimination, in a number
of steps bound by a \emph{polynomial} in the dimension $|\Pi|$ of
$\Pi$. On the other side, the system must be \emph{complete}:
every P-Time Turing machine can be encoded, and simulated
by means of a derivation.

We prove correctness by introducing the proof nets for the sequent
calculus in Figure~\ref{figure:ILAL}. 
Proof nets are the right syntax for calculating a computational
complexity because their computational steps are truly primitive,
and close to pointer manipulations, performed by real machines.
Every step is a (graphical) re-wiring of links, whose cost can be
fairly taken as a unit.

The proof of completeness rests on the 
definition of a concrete syntax for the derivations.
This choice is due to the need of readability. The use of the derivations 
of the sequent calculus are not very comfortable as a programming language.
Proof nets would be OK, but very cumbersome in terms of space, and not
everybody is akin to use them to program. 

\section{Proof Nets}
\label{section:Proof Nets}

\begin{figure}
\centering\epsfig{file=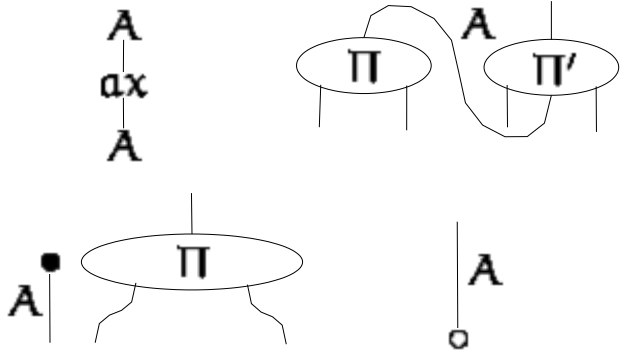,
bbllx=0, bblly=450, bburx=235, bbury=596,clip=}
\caption{PNs of ILAL: axiom, cut, weakening, and unit}
\label{figure:net-def-s}
\end{figure}

\begin{figure}
\centering\epsfig{file=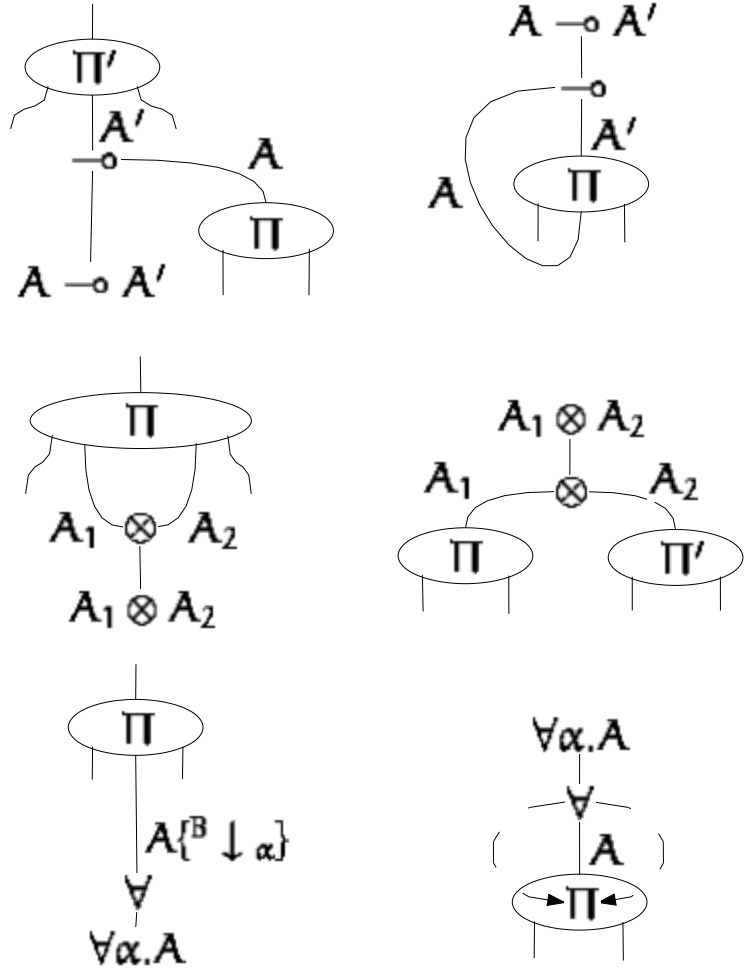,
bbllx=0, bblly=220, bburx=295, bbury=595,clip=}
\caption{PNs of ILAL: the second order and multiplicative fragment}
\label{figure:net-def-l}
\end{figure}

\begin{figure}
\centering\epsfig{file=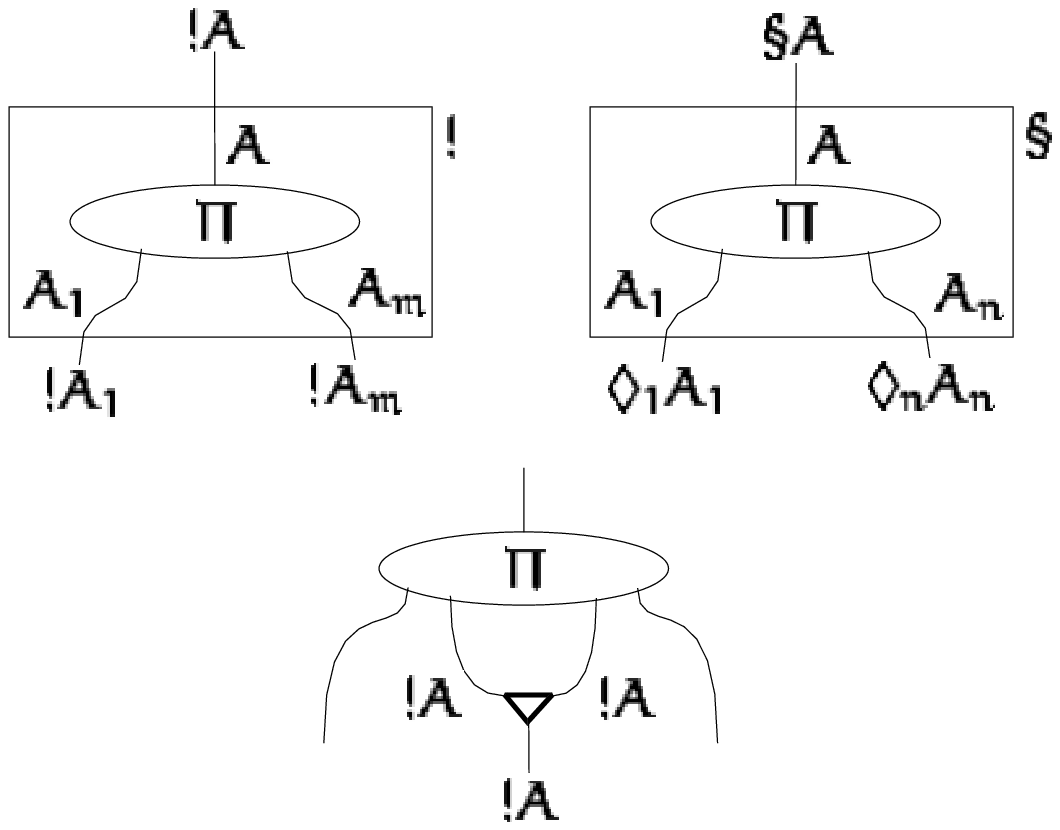,
bbllx=0, bblly=340, bburx=322, bbury=596, clip=}
\caption{PNs of ILAL: the polynomial fragment}
\label{figure:net-def-b}
\end{figure}

The Proof Nets (PNs) for ILAL
are the graphs in Figure~\ref{figure:net-def-s},
\ref{figure:net-def-l}, 
and~\ref{figure:net-def-b}.
The PNs have a single output, and as many inputs as needed, possibly none.
The output, also called root, is the link on top of the graph.
The inputs, also called assumptions, are all the other links.

Figure~\ref{figure:net-def-s} introduces the axiom , the cut, the
weakening and the unit.
The axiom, labeled $ax$, is a PN with a single input and a single output.
If $\Pi$ and $\Pi'$ are two PNs, the first with its output labeled 
by $A$, and the second with an input labeled by $A$, then
the graph obtained by
plugging the output of $\Pi$ into the input of $\Pi'$ is a PN.
With more traditional terminology, this is cutting the conclusion of $\Pi$
with the assumption of $\Pi'$.
Take again a PN $\Pi$. By putting a wire with a single input and no
conclusions at all aside $\Pi$ yields a new PN: this is traditional weakening.
Observe that the new, fake assumption is labeled by any formula $A$,
namely, unlike traditional Linear logic, ILAL has an \emph{unconstrained}
weakening. 
Finally, the unit.
It has a conclusion, but no inputs, like Linear
logic's unit $\mathbf{1}$. However, any formula can label our unit,
and not only $\mathbf{1}$. 
Our unit serves to close the set of PNs with respect to
the cut elimination, in
presence of the unconstrained weakening.

Figure~\ref{figure:net-def-l} defines the PNs for the second
order and multiplicative fragment of ILAL.
Everything is quite standard.
Assume $\Pi$ and $\Pi'$ be two PNs.
Then, a new PN is obtained by wiring the conclusions/assumptions
of $\Pi$/$\Pi'$ as depicted.
The \emph{introduction of a new root} in the proof nets stands for an 
\emph{introduction to the right} in sequent calculus terminology,
while a \emph{new input} is like an
\emph{introduction to the left} of the sequent calculus.
Notice the $\forall$-introduction to the right (the lower-rightmost PN)
in Figure~\ref{figure:net-def-l}. 
Its dashed links must point to all the wires
of $\Pi$ whose labeling formula has $\alpha$ among its free variables.
Moreover, no input wire of $\Pi$ 
must be pointed by the dashed links. This is like 
the usual $\forall$-introduction to the right: it requires that the
variable being universally quantified is not a free variables of the
assumptions. 
Our $\forall$-introduction to the right is not like in standard PNs
of Linear logic. The standard construction, by means of a box, introduces an
artificial sequentialization in the construction of the PNs that requires
the use of commuting conversions to get the cut elimination. Our construction
has not this drawback,
simplifying the estimation of the cut elimination complexity. 

Figure~\ref{figure:net-def-b} defines the PNs for the polynomial
fragment of ILAL. Assume $\Pi$ be a PN. A new PN is obtained
either by  enclosing $\Pi$ into a box, or by contracting two of its inputs,
labeled by a \emph{modal} formula $!A$,
into a single input, labeled by $!A$ as well.
There are two kinds of boxes.
Any $!$-box has \emph{at most} one input, labeled by a $!$-modal formula. 
So, in Figure~\ref{figure:net-def-b}, $m\leq 1$.
On the contrary, there are not restrictions on the inputs of the
$\S$-box: every $\lozenge_i$ belongs to $\{!, \S\}$, and $n$ is any integer,
possibly $0$. The big difference between the two boxes will be appreciated
when defining the cut elimination: a $!$-box can be duplicated, but every
$\S$-box cannot.

\section{Cut Elimination on Proof Nets}
\label{section:Cut Elimination on Proof Nets}

\begin{figure}[htbp]
\centering\epsfig{file=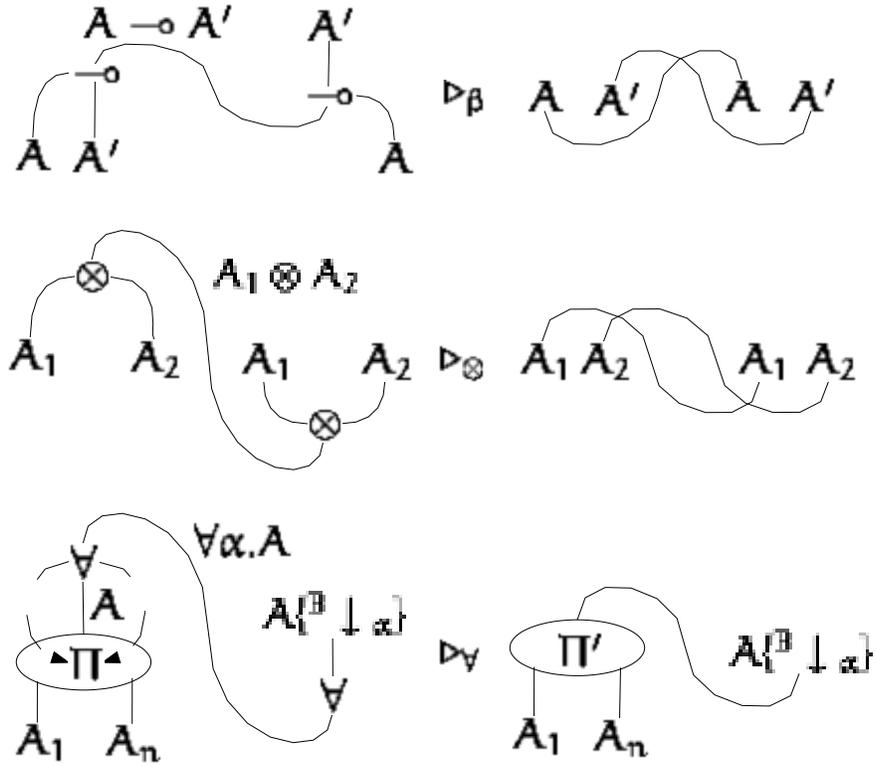,
bbllx=0, bblly=299, bburx=333, bbury=596, clip=}
\caption{Cut elimination: the linear steps}
\label{figure:red-steps-l-base}
\end{figure}

\begin{figure}[htbp]
\centering\epsfig{file=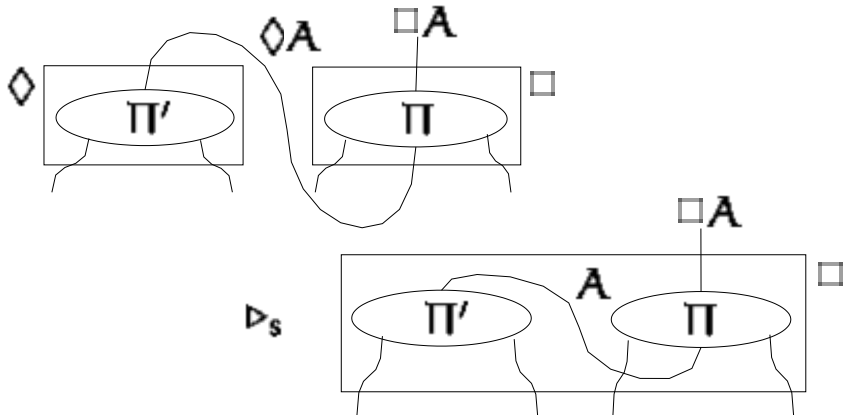,
bbllx=0, bblly=435, bburx=432, bbury=596, clip=}
\caption{Cut elimination: the shifting step}
\label{figure:red-steps-s}
\end{figure}

\begin{figure}[htbp]
\centering\epsfig{file=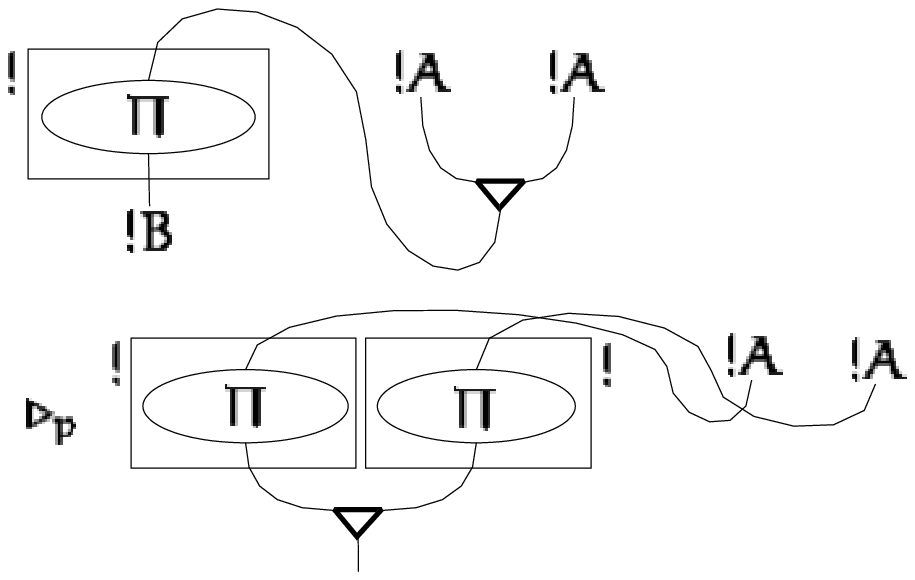,
bbllx=0, bblly=425, bburx=275, bbury=596, clip=}
\caption{Cut elimination: the polynomial step}
\label{figure:red-steps-p}
\end{figure}

\begin{figure}[htbp]
\centering\epsfig{file=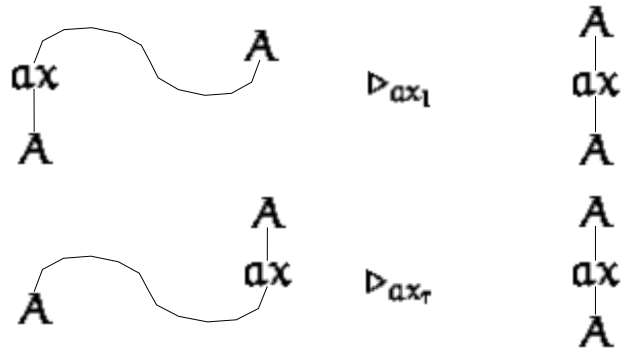,
bbllx=0, bblly=450, bburx=333, bbury=596, clip=}
\caption{Cut elimination: the garbage collection generated by the
axioms}
\label{figure:red-steps-gcax}
\end{figure}

\begin{figure}[htbp]
\centering\epsfig{file=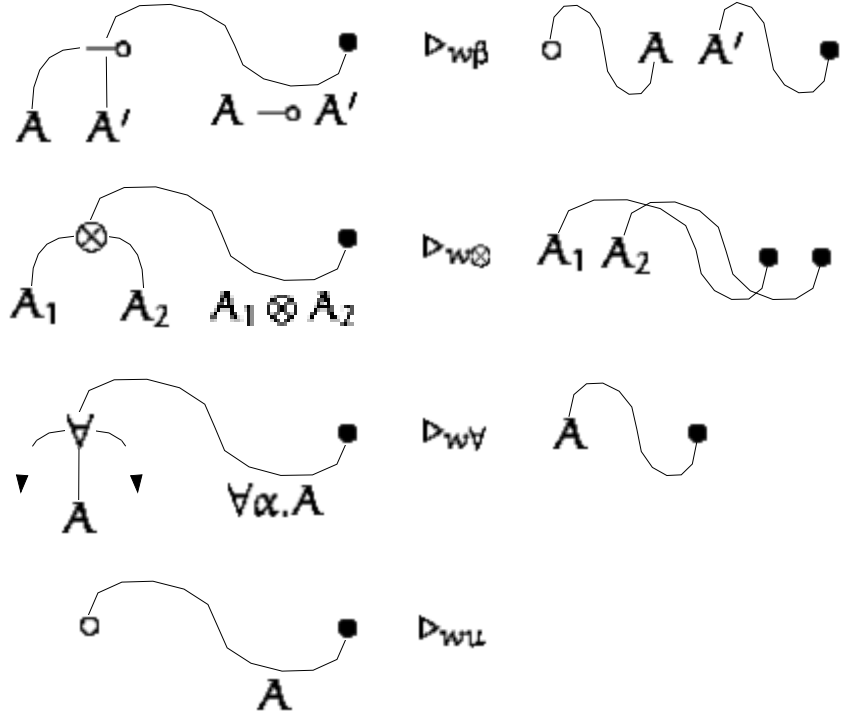,
bbllx=0, bblly=317, bburx=333, bbury=596, clip=}
\caption{Cut elimination: the garbage collection generated by the
unconstrained weakening}
\label{figure:red-steps-gcw}
\end{figure}

\begin{figure}[htbp]
\centering\epsfig{file=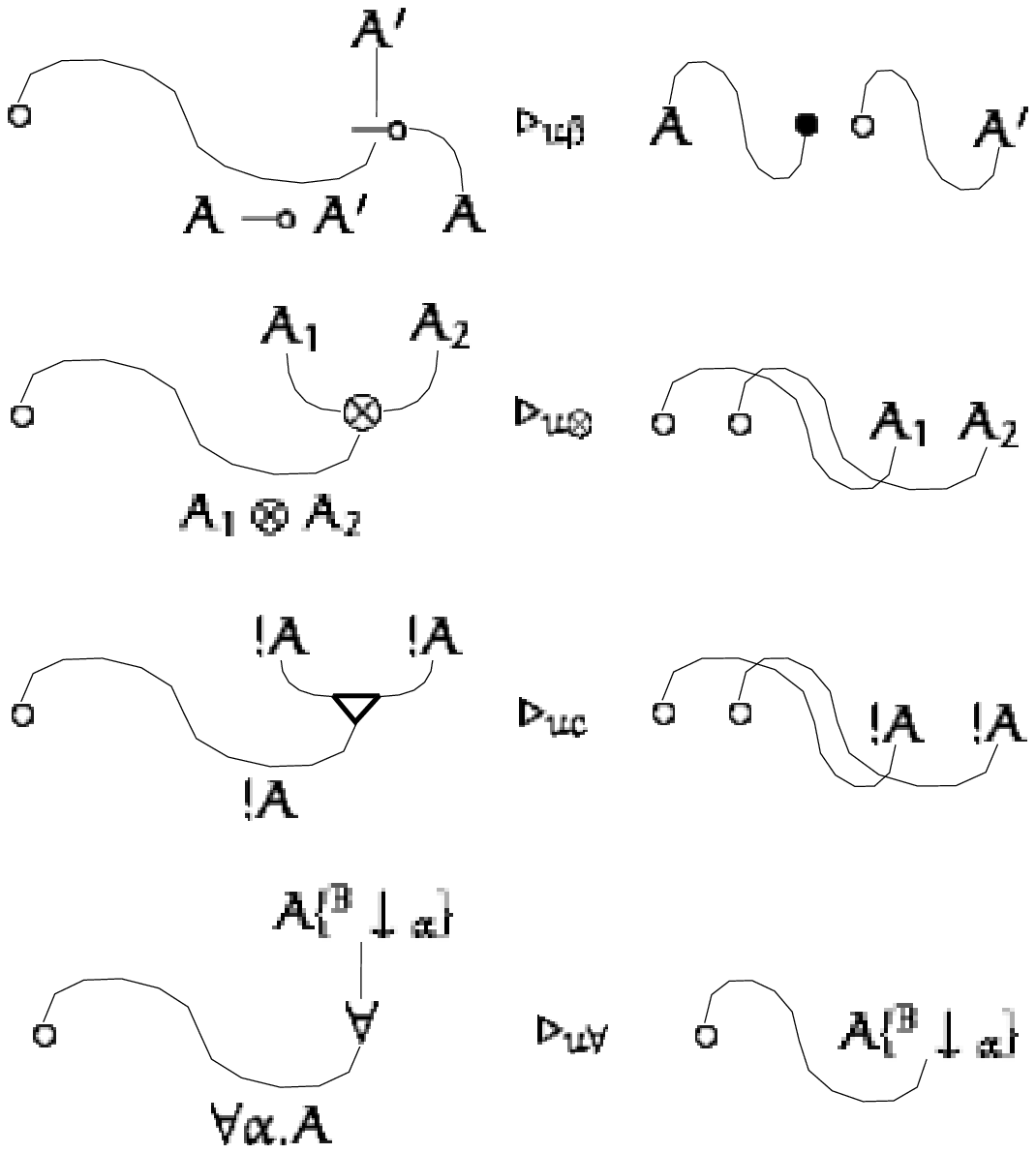,
bbllx=0, bblly=245, bburx=320, bbury=596, clip=}
\caption{Cut elimination: the garbage collection generated by the unit}
\label{figure:red-steps-gcu}
\end{figure}

\begin{figure}[htbp]
\centering\epsfig{file=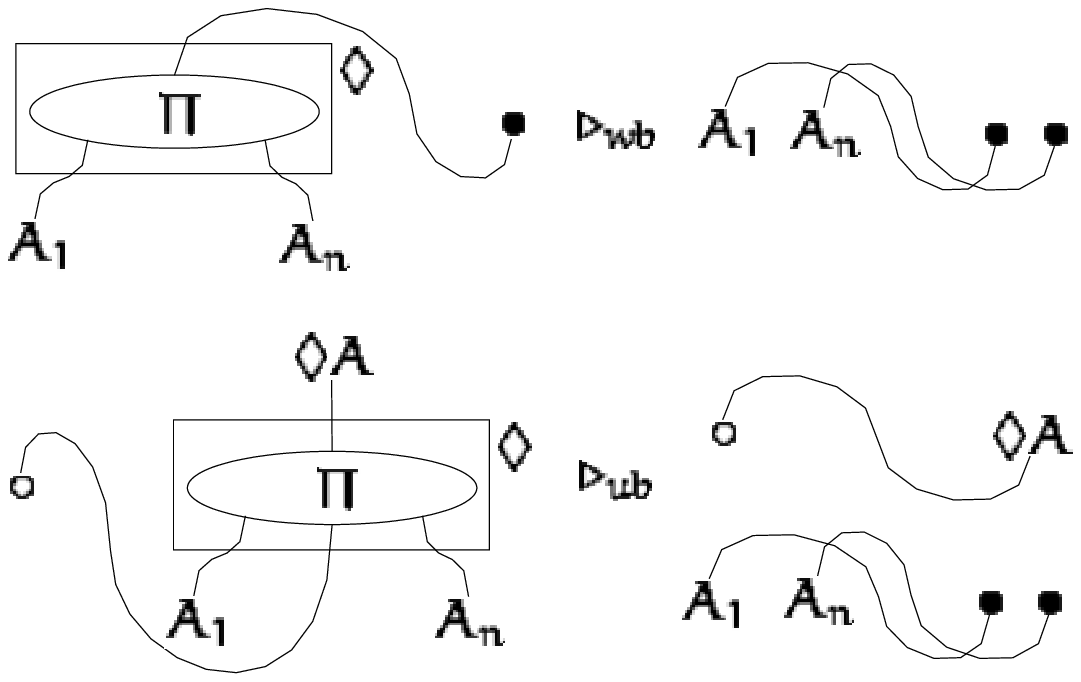,
bbllx=0, bblly=395, bburx=318, bbury=596, clip=}
\caption{Cut elimination: the garbage collection generated erasing a box}
\label{figure:red-steps-gcb}
\end{figure}

The main rules for eliminating the cuts are in 
Figures~\ref{figure:red-steps-l-base},
\ref{figure:red-steps-s},
and
\ref{figure:red-steps-p}.
Figure~\ref{figure:red-steps-gcax},
\ref{figure:red-steps-gcw},
\ref{figure:red-steps-gcu}, and
\ref{figure:red-steps-gcb}
complete the cut elimination with garbage collection steps.
The cut elimination rewrites
graphs into other graphs which are not necessarily Proof nets of
ILAL, but this will not be armuful.

Figure~\ref{figure:red-steps-l-base},
introduce the \emph{linear steps}.
Figure~\ref{figure:red-steps-s} introduces the \emph{shifting step},
and Figure~\ref{figure:red-steps-s} the \emph{polynomial step}.
This terminology is related to the cost of eliminating the
corresponding cuts. The garbage collection cost will not be accounted
because its steps only destroy existing structure:
this means that the cost will never be greater than
the dimension of the net being reduced.

Figure~\ref{figure:red-steps-l-base} 
defines the \emph{linear} cut elimination
$\rewl =
 \rewrel_{\beta}  \cup
 \rewrel_{\otimes}\cup
 \rewrel_{\forall}$.
The steps $\rewrel_{\beta}$ and $\rewrel_{\otimes}$ 
describe how a pair of $\linear$ or $\otimes$-nodes annihilate each other.
The step $\rewrel_{\forall}$ annihilates two $\forall$-nodes and produces
$\Pi'$ from $\Pi$ by substituting $B$ for every free occurrence of
$\alpha$ in the formulas that label the edges of $\Pi$, pointed to by the
dashed links.

Figure~\ref{figure:red-steps-s}
defines the shifting cut elimination step 
$\rews$, which shifts a net $\Pi'$, contained in a box,
into another box.
The $\Box$-box can be either a $\S$-box, or a $!$-box.

Figure~\ref{figure:red-steps-p} defines
the rewriting relation $\rewrel_{p}$. It \emph{only} duplicates $!$-boxes.

Figure~\ref{figure:red-steps-gcax}
the set of steps that compresses a sequence axiom/cut into a
single axiom.

Figure~\ref{figure:red-steps-gcw}
defines a second set of \emph{garbage collecting} cut elimination steps.
The use of the unconstrained weakening requires to consider all the possible
configurations where the conclusion of some (sub-)net is plugged into
the fake input of a weakening node. In such a case, the cut elimination
proceeds just by erasing structure. In particular, for preserving the
structural invariance that a cut link plugs the conclusion of a (sub-)net 
into the assumption of another (sub-)net, $\rewrel_{w\beta}$ introduces 
the unit net to erase the nodes of which the left link of
the $\linear$-node is an input.
Figure~\ref{figure:red-steps-gcu} 
defines the \emph{garbage collecting} cut elimination steps relative to our unit.

Finally, Figure~\ref{figure:red-steps-gcb}
shows what happens when erasing a box, using either a weakening or a unit.
In particular, notice that $\rewrel_{wu}$ erases a box from the bottom: 
so the unit keeps
erasing from $\lozenge A$ upward, while the weakening go downward.

We call $\rewl$-\emph{normal} a net $\Pi$ without $\rewl$-redexes.
We shall also use the analogous terminology for $\rews$ and $\rewp$.
If a net does not contain redexes of $\rewrel \equiv \rewl \cup \rews \cup
\rewp$ it is simply \emph{normal}.
Of course, a net can also be garbage collected, so it is normal with respect to
the rules in Figure~\ref{figure:red-steps-gcax} through~\ref{figure:red-steps-gcb}.
However we shall not pay very much attention to the garbage collection, when
concerned to the complexity of the cut elimination.
Ideed, the garbage collection can be ``runned'' at any instant without
significant overhead: it strictly decreases the amount of existing structure.

\subsection{Properties of the cut elimination}
We observed that $\rewrel$ rewrites graphs into graphs and not Proof nets into
Proof nets. This is not a problem: 
\begin{proposition}
The set of Proof nets for ILAL is closed under $\rewrel$. 
\end{proposition}

This can be proved in few steps.
The Proof nets of ILAL, \emph{without units and weakenings},
can be embedded into those of \emph{functorial}
ILL, whose characterizing rules are recalled in
Figure~\ref{figure:functorial-ILL}.
\begin{figure}
\begin{eqnarray*}
&
\dedrule{(!)}
{A_1,\ldots,A_n\vdash B}
{!A_1,\ldots,!A_n\vdash !B}
&\\
&
\dedrule{(Dereliction)}
{\Gamma, A\vdash B}
{\Gamma, !A\vdash B}
&\\
&
\dedrule{(Digging)}
{\Gamma, !!A\vdash B}
{\Gamma, !A\vdash B}
&
\end{eqnarray*}
\caption{Functorial ILL: characterizing rules}
\label{figure:functorial-ILL}
\end{figure}
The only point worth specifying on the embedding is that
it maps every occurrence of $\S$ into an occurrence of $!$;
the rest is a one-one correspondence.
The closure extends to the whole language of ILAL Proof nets for some
simple reasons.
One of the two nets involved in the garbage collecting cut eliminations
is always an \emph{unconnected} component: either a unit or a weakening.
Unit does not have inputs, so it does not create any problems concerning the
construction order inherent to an inductive definition: given any net $\Pi$,
we can always take a unit and cut its conclusion with any assumption of $\Pi$,
with compatible type. Weakenings behave almost analogously. A weakening is always
associated to some well formed net $\Pi$. 
Suppose that the elimination of a cut between a weakening and the root of a net $\Pi'$
yields new cuts between the roots of the sub-nets of $\Pi'$ and some weakenings.
Then the newly generated weakenings can be thought of as introduced in association
with $\Pi$ itself.

The Proof nets of ILAL are also a good computational language:

\begin{proposition}
$\rewrel$ is Church-Rosser.
\end{proposition}

Start, again, from the Proof nets of ILAL, \emph{without units and weakenings},
and embed them into those of \emph{functorial} ILL.
The strong normalizability of \emph{functorial} ILL\ implies the same property for
the considered fragment of ILAL.
As we alrady observed,
the garbage collection certainly does not break 
the strong normalizability, because it strictly decrease the size of the nets.
Now, to check that Church-Rosser holds, just verify that the few critical
pairs of $\rewrel$ are confluent. By the way, the critical pairs are the same as
those of the Proof nets for (functorial) ILL.

\section{P-Time Correctness}
\label{section:P-Time Correctness}

P-Time correctness means that, for any proof net $\Pi$, the number
of cut links that must be eliminated to get to the normal form
of $\Pi$ is bound by a polynomial in the dimension of $\Pi$.

This is the statement we shall prove by the end of this section.

It will turn out that the bound is:
\begin{eqnarray*}
O(\maximaldimension^{3^{\maximaldepth}}(\Pi))
\enspace,
\end{eqnarray*}
where 
$\maximaldimension(\Pi)$ is the dimension of $\Pi$,
and $\maximaldepth$ is the maximal depth of $\Pi$.
The dimension is, essentially, the number of nodes in $\Pi$.
The depth of $\Pi$ is a purely structural property of $\Pi$, and will be
introduced in a few.

The main tool to develop the proof of P-Time
correctness is to find a measure that describes how 
$\maximaldimension(\Pi)$ changes, as the cut elimination proceeds.
Indeed, the number of nodes in a net always bound the number of the
cut links that can be eliminated.

\subsection{Proving P-Time Correctness}

Every net can be stratified in \emph{levels}:

\begin{definition}[Level of a net]
For any net $\Pi$, a node of $\Pi$ is at \emph{level} $l$
if it is enclosed into $l$ boxes of kind $!$ and/or $\S$.

The maximal depth of $\Pi$ is $\maximaldepth(\Pi)$, or simply
$\maximaldepth$, if no ambiguity can exist.
\end{definition}
\noindent

\begin{definition}[Dimensions of a Net]
Let $\Pi$ be a net, and $l\leq\maximaldepth$.
\begin{itemize}
\item
The \emph{dimension} $\dimlev_l(\Pi)$ of $\Pi$
\emph{at level} $l$ is the number of $\forall$,
contraction nodes, $!$-boxes, and $\S$-boxes, plus
$\linear$ and  $\otimes$-nodes, introduced either to the left, or to the right,
at level $l$.
\item
The \emph{level-by-level} dimension of $\Pi$ is:
\begin{eqnarray*}
\lblmeasure(\Pi)&=&
\langle
\dimlev_{0}(\Pi),
\ldots,
\dimlev_{i}(\Pi),
\ldots,
\dimlev_{\maximaldepth}(\Pi)
\rangle
\end{eqnarray*}
\item
The \emph{maximal dimension} $\maximaldimension(\Pi)$ is simply
$\sum_{l=0}^{\maximaldepth} \dimlev_l(\Pi)$.
\end{itemize}
Of course, when there are not ambiguities, the argument $\Pi$ is omitted.
\end{definition}
\noindent
\begin{remark}
\begin{itemize}
\item
The nodes at the same level $l$ can be ``spread'' in various boxes, each
contributing to form the level $l$.
\item
The space of tuples which $\lblmeasure$ belongs to is a well founded
order, under the \emph{lexicographic} relation $\succeq$. In particular,
$\succ$ is the non reflexive part of $\succeq$.
\end{itemize}
\end{remark}

\begin{figure}[htbp]
\centering\epsfig{file=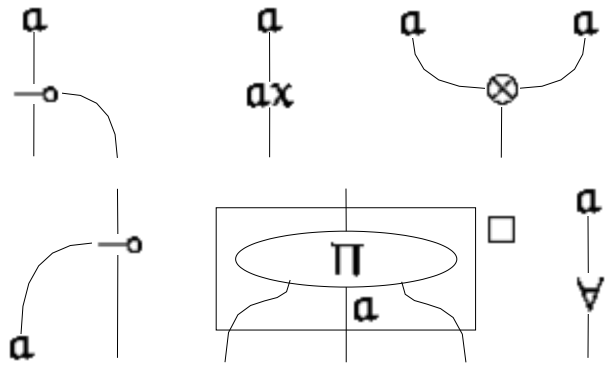,
bbllx=0, bblly=453, bburx=300, bbury=596, clip=}
\caption{The weight of the sub-nets: first case}
\label{figure:weight-for-sub-nets-1st-case}
\end{figure}

\begin{figure}[htbp]
\centering\epsfig{file=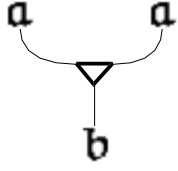,
bbllx=0, bblly=525, bburx=70, bbury=596, clip=}
\caption{The weight of the sub-nets: second case}
\label{figure:weight-for-sub-nets-2nd-case}
\end{figure}

\begin{figure}[htbp]
\centering\epsfig{file=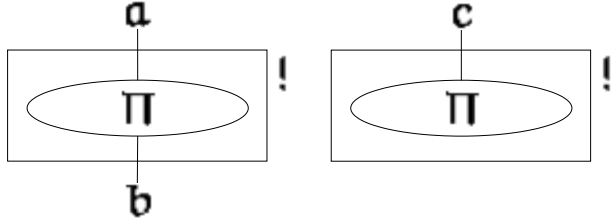,
bbllx=0, bblly=505, bburx=240, bbury=596, clip=}
\caption{The weight of the sub-nets: third case}
\label{figure:weight-for-sub-nets-3rd-case}
\end{figure}

Every point of a given net $\Pi$ can be taken as the root of
a \emph{weighted} sub-net:

\begin{definition}[Weight of a Net]
Let $\Pi$ be a net.
The weight $\weight(\Pi)$ of $\Pi$ is a partial function from
points of $\Pi$ to integers.
If $a$ is any point on a link of $\Pi$:
\begin{eqnarray*}
\weight(\Pi)(a) &=& 0 
 \text{ with $a$ as in Figure~\ref{figure:weight-for-sub-nets-1st-case}, where}
\\
&&\phantom{ 0 }
\text{$\Box\in\{!,\S\}$, and the $ax$-link is an input of $\Pi$}\\
\weight(\Pi)(a) &=& \weight(\Pi)(b)
 \text{ with $a$, and $b$ as in Figure~\ref{figure:weight-for-sub-nets-2nd-case}}\\
\weight(\Pi)(a) &=& \weight(\Pi)(b) + 1
 \text{ with $a$, and $b$ as in Figure~\ref{figure:weight-for-sub-nets-3rd-case}}\\
\weight(\Pi)(a) &=& 1
 \text{ with $c$ as in Figure~\ref{figure:weight-for-sub-nets-3rd-case}}
\end{eqnarray*}
$\weight(\Pi)$ is undefined on any other point.
\end{definition}

\begin{definition}[Weight of a Contraction]
The weight $\weight(\nabla)$ of any instance $\nabla$ of a contraction node 
in a net $\Pi$ is $\weight(\Pi)(a)$ if $a$ labels the input of $\nabla$.
\end{definition}
\begin{remark}
\begin{itemize}
\item
$\weight(\nabla)$ is the number of $!$-boxes that can be duplicated by
$\nabla$, and that are at the same level as $\nabla$ is at;
\item
the points whose $\weight$ is $0$ are those where a contraction node stops
moving down, through a net, during the cut elimination;
\item
every contraction node is as ``heavy'' as the weight
of the net rooted at its input;
\item
last, but not at all least, $\weight$ is finite at every level, because the nets
are defined inductively, and the cut elimination preserves their inductive
structure.
\end{itemize}
\end{remark}

\begin{definition}[``Refined'' Dimension of a Net]
\label{definition:refined-dimension}
Let $\Pi$ be a given net, and $l$ any integer of $\mathbb{Z}$.
\begin{itemize}
\item
$n_l(\Pi)$ is the number of nodes $\forall$ plus
the $\linear$, and the $\otimes$-nodes, introduced either to the left, or to the right, at level $l$ in $\Pi$;
\item
$c^w_l(\Pi)$ is the number of contraction nodes at $l$ in $\Pi$
with weight $w$;
\item
$b_l(\Pi)$ is the total number of $!$-boxes, and $\S$-boxes at $l$
in $\Pi$;
\item
$\maximalweight_l(\Pi)$ is the maximal weight of the contraction nodes
at $l$ in $\Pi$.
\end{itemize}
In particular, each of the quantities here above can assume any value in
$\mathbb{N}$ if $0\leq l\leq \maximaldepth(\Pi)$. Otherwise, their value can
only be $0$.

When clear from the context, we omit $\Pi$, and also the level $l$. 
\end{definition}

The complexity bound follows from using a specific reduction strategy.
The next definitions, and lemmas will serve to introducing
such a strategy.

\begin{definition}[Normalizing Measure of a Net]
For any net $\Pi$, its \emph{cut measure} at level 
$0\leq l\leq \maximaldepth$ is:
\begin{eqnarray*}
\cutmeasure_l(\Pi)&=&
\langle
c^{\maximalweight_{l-1}}_{l-1},
\ldots,
c^1_{l-1},
b_{l-1},
n_{l}
\rangle
\end{eqnarray*}
\end{definition}

\begin{remark}
\begin{itemize}
\item
The measure involves \emph{two} levels of the net.
If $l=0$, by Definition~\ref{definition:refined-dimension}, the
rightmost component can assume any natural value. All the others are $0$.
\item
The space of tuples which $\cutmeasure$ belongs to is a well founded
order, under the \emph{lexicographic} relation $\succeq$.
\end{itemize}
\end{remark}

\begin{definition}[$l$-normal Net]
Let $\Pi$ be a net, and $l\leq\maximaldepth$.
We say that $\Pi$ is $l$-normal when:
\begin{itemize}
\item
$\Pi$ is $\rewl$-normal at every level $0\leq i \leq l$, and
\item
$\Pi$ is $\rews\cup\rewp$-normal at all levels $0\leq i \leq l-1$.
\end{itemize}
\end{definition}

\begin{fact}
\label{fact:maximaldepth}
If $\Pi$ is $\maximaldepth$-normal, then $\Pi$ is, in fact, \emph{normal}.
\end{fact}
$\maximaldepth$-normality means no $\rewl$-redexes at
any level, and no $\rews\cup\rewp$-redexes at all levels but $\maximaldepth$.
Assuming the existence of a $\rews\cup\rewp$-redex at $\maximaldepth$,
means to have some $!$ or $\S$-box at $\maximaldepth$, against the definition
of $\maximaldepth$ for $\Pi$.

\begin{fact}
\label{fact:ls-decreasing-l}
Let $\Pi$ be $l-1$-normal, with $l\leq\maximaldepth$,
and such that $\Pi\rewl\Pi'$ by reducing a redex at $l$.
Then:
\begin{eqnarray}
\cutmeasure_l(\Pi)=\langle \ldots,n_{l}\rangle
&\succ&
\langle \ldots,n_{l}-1\rangle=\cutmeasure_l(\Pi')
\label{eqn:ls-decreasing-l1}
\\
\lblmeasure(\Pi)=\langle \dots,\dimlev_{l},\ldots\rangle
&\succ&
\langle \dots,\dimlev_{l}-1,\ldots\rangle=\lblmeasure(\Pi')
\label{eqn:ls-decreasing-l2}
\end{eqnarray}
\end{fact}
\noindent
\eqref{eqn:ls-decreasing-l1} holds because
the reduction of a $\rewl$-redex at level $l$ erases one node
among $\forall, \linear, \otimes$.
So, \eqref{eqn:ls-decreasing-l2}
simply follows from \eqref{eqn:ls-decreasing-l1}.

\begin{fact}
\label{fact:ls-decreasing-s}
Let $\Pi$ be $l-1$-normal, with $l\leq\maximaldepth$,
and such that $\Pi\rews\Pi'$ by reducing a redex at $l-1$.
Then:
\begin{eqnarray}
\label{eqn:ls-decreasing-s1}
\lefteqn{
\cutmeasure_l(\Pi)=\langle \ldots,b_{l-1}(\Pi),n_{l}(\Pi)\rangle
\succ
}\\
&&
\qquad
\qquad
\langle \ldots,b_{l-1}(\Pi)-1,n_{l}(\Pi)\rangle=\cutmeasure_l(\Pi')
\nonumber
\\
\label{eqn:ls-decreasing-s2}
\lefteqn{
\lblmeasure(\Pi)=
\langle \ldots,\dimlev_{l-1}(\Pi),
               \dimlev_{l}(\Pi),
	\ldots
\rangle
\succ
}\\
&&
\qquad
\qquad
\langle \ldots,\dimlev_{l-1}(\Pi)-1,
               \dimlev_{l}(\Pi),
	\ldots \rangle=
\lblmeasure(\Pi')
\nonumber
\enspace .
\end{eqnarray}

\end{fact}
\noindent
Fact~\ref{fact:ls-decreasing-s} is obvious for
the reduction merges the border of two boxes, so decreasing their 
number at $l-1$.

\begin{fact}
\label{fact:ls-decreasing-p}
Let $\Pi$ be $l-1$-normal, with $l\leq\maximaldepth$,
and such that $\Pi\rewp\Pi'$ by eliminating a cut at $l-1$, which
involves a contraction node $\nabla$
with weight $w\leq\maximalweight_{l-1}(\Pi)$.
Then:
\begin{eqnarray}
\cutmeasure_l(\Pi)
&=&
\langle \ldots,c^w_{l-1}(\Pi),
               c^{w-1}_{l-1}(\Pi),
	 \ldots,b_{l-1}(\Pi),n_l(\Pi)
\rangle\\
&\succ&
\langle \ldots,c^w_{l-1}(\Pi)-1,
               c^{w-1}_{l-1}(\Pi'),
	\ldots,b_{l-1}(\Pi'),
	       n_l(\Pi')
\rangle\nonumber\\
&=&
\cutmeasure_l(\Pi')
\nonumber
\\
\lblmeasure(\Pi)
&=&
\langle \ldots,\dimlev_{l-1}(\Pi),
               \dimlev_{l}(\Pi),
	\ldots,\dimlev_{\maximaldepth}(\Pi),
\rangle
\\
&\prec&
\langle \ldots,\dimlev_{l-1}(\Pi'),
               \dimlev_{l}(\Pi'),
	\ldots,\dimlev_{\maximaldepth}(\Pi'),
\rangle
\nonumber
\\	
&=&
\lblmeasure(\Pi')
\nonumber
\end{eqnarray}
where:
\begin{eqnarray}
c^{w-1}_{l-1}(\Pi')&\leq& c^{w-1}_{l-1}(\Pi)+1
\label{eqn:one-step-reductp-1}\\
b_{l-1}(\Pi')&=& b_{l-1}(\Pi)+ 1
\label{eqn:one-step-reductp-2}\\
n_{l}(\Pi')&\leq& n_{l}(\Pi)+ n_{l}(\Pi)
\label{eqn:one-step-reductp-3}\\
\dimlev_{l-1}(\Pi')
       &=&\dimlev_{l-1}(\Pi)+1
\label{eqn:one-step-reductp-4}\\
\dimlev_{l}(\Pi')
       &\leq&\dimlev_{l}(\Pi)+
             \dimlev_{l}(\Pi)
\label{eqn:one-step-reductp-5}
\end{eqnarray}
where $l\leq i\leq\maximaldepth$.
\end{fact}
\noindent
\eqref{eqn:one-step-reductp-1} holds because $\nabla$ may be propagated
below the just duplicated bos. In such a case, the weight decreases by one.
\eqref{eqn:one-step-reductp-2} holds because the duplication 
introduces a $!$-box more than those in $\Pi$ at $l-1$.
From this, it is obvious \eqref{eqn:one-step-reductp-4} as well.
\eqref{eqn:one-step-reductp-3} holds because the introduction of a new
$!$-box at level $l-1$ means to make one copy of \emph{at most} all the
nodes $\forall, \linear$, and $\otimes$
at level $l$ of $\Pi$.
This gives meaning also to \eqref{eqn:one-step-reductp-5}.

\begin{proposition}
\label{proposition:SNSC-rewp}
Let $\Pi$ be $l-1$-normal, where $l\leq\maximaldepth$.
Assume that $\Pi$ rewrites to $\Pi'$ by eliminating \emph{all} 
$\rewp$-cuts at level at $l-1$. Then, 
$\rewp$ is strongly normalizable, and strongly confluent.
\end{proposition}
Strong normalizability trivially follows from
Fact~\ref{fact:ls-decreasing-p}.
Strong confluence follows from the absence of critical pairs in $\rewp$.

\begin{proposition}
\label{proposition:SNSC-rewl-rews}
Let $\Pi$ be $l-1$-normal, without $\rewp$-redexes at level $l-1$, where
$l\leq\maximaldepth$.
Assume that $\Pi$ rewrites to $\Pi'$ by eliminating \emph{all} 
$\rews$-cuts at level at $l-1$, and all $\rewl$-redexes at level $l$,
without assuming any precedence among the $\rews\cup\rewl$-redexes.
Then, 
$\rews\cup\rewl$ is strongly normalizable, and strongly confluent.
\end{proposition}
Strong normalizability follows from
Fact~\ref{fact:ls-decreasing-l}, and~\ref{fact:ls-decreasing-s}.
Both imply that $\cutmeasure$, and $\lblmeasure$ have a common upper bound
as the elimination of $\rews\cup\rewl$-redexes proceeds.
Strong confluence follows from the absence of critical pairs in
$\rews\cup\rewl$.

The two, just given, properties support the definition of a reduction strategy:

\begin{definition}[Cut Elimination Strategy]
Let $\Pi$ be $l-1$-normal.
The cut elimination strategy $\strategy$ reduces redexes of
$\rewl\cup\rews\cup\rewp$ in the following order:
\begin{itemize}
\item
firstly, \emph{all} the $\rewp$-redexes at $l-1$,
\item
secondly, \emph{all} the $\rews$-redexes at $l-1$,
and the $\rewl$-redexes at $l$, in any order.
\end{itemize}
Then, $\strategy$ stops.
\end{definition}

\begin{proposition}
\label{proposition:basic-complexity}
Let $\Pi$, and $\Pi'$ be such that $\Pi$ is $l-1$-normal,
 and $\Pi\strategy\Pi'$.
Then:
\begin{enumerate}
\item
$\Pi'$ is $l$-normal;
\item
$\Pi\strategy\Pi'$ takes at most
$6\cdot \maximaldimension^3(\Pi)$ steps.
\item
$\dimlev_i(\Pi') \leq 6\cdot\maximaldimension^3(\Pi)$, 
for all $l\leq i\leq\maximaldepth$.
\end{enumerate}
\end{proposition}
\begin{proof}
The first point is true  by definition of $\strategy$.

Let us focus on the second point.
Assume that:
\begin{eqnarray*}
\cutmeasure_l(\Pi)
&=&
\langle
\overbrace{
\dimlev_{l-1},0,\ldots,0
}^{\maximalweight_{l-1}=\dimlev_{l-1}\text{ times}},
\dimlev_{l-1},
\dimlev_{l}
\rangle
\enspace .
\end{eqnarray*}
$\cutmeasure_l(\Pi)$ here above is the worst possible assumption 
with respect of the number of cut elimination steps, necessary to normalize
$\Pi$ at level $l$, because:
\begin{itemize}
\item
we assume that all the contraction nodes at $l-1$,
\ie\ as many as $\dimlev_{l-1}(\Pi)$, have maximal
weight. We saw that the weight of a contraction node $\nabla$ is the
maximal number of $!$-boxes at $l-1$ that $\nabla$ can duplicate.
Forcefully, the $!$-boxes at $l-1$ can not be more than
$\dimlev_{l-1}(\Pi)$. This defines as many leftmost components 
of $\cutmeasure_l(\Pi)$ as $\dimlev_{l-1}(\Pi)$;
\item
we assume to have as many $!$/$\S$-boxes as
possible at $l-1$, namely $\dimlev_{l-1}(\Pi)$, defining the
second component of $\cutmeasure_l(\Pi)$ from its right;
\item
we assume to have as many nodes as possible at $l$ in $\Pi$,
namely $\dimlev_{l}(\Pi)$, defining the
rightmost component of $\cutmeasure_l(\Pi)$.
\end{itemize}
Then, we make the hypothesis that every contraction node, $!$-box,
$\S$-box at $l-1$, and every node at $l$ contributes to form a redex.
Finally, we apply $\strategy$, and we observe the behavior of
$\cutmeasure_l(\Pi)$:
\begin{eqnarray}
\label{eqn:basic-round}
\lefteqn{\cutmeasure_l(\Pi)}
\nonumber\\
&\equiv&
\langle
\dimlev_{l-1},0,\ldots,0,
\dimlev_{l-1},
\dimlev_{l}
\rangle\nonumber\\
&&
\ldots\text{after $1$ step of $\rewp$, from $\Pi$}\ldots\nonumber\\
&\succ&
\langle
\dimlev_{l-1}-1,1,\ldots,0,
\dimlev_{l-1}+1,
\dimlev_{l}+\dimlev_{l}
\rangle\nonumber\\
&&
\ldots\text{after $i$ steps of $\rewp$, from $\Pi$}\ldots\nonumber\\
&\succ&
\langle
\dimlev_{l-1}-i,i,\ldots,0,
\dimlev_{l-1}+i,
\dimlev_{l}+i\cdot\dimlev_{l}
\rangle\nonumber\\
&&
\ldots
\text{after $\dimlev_{l-1}$ steps of $\rewp$, from $\Pi$}
\ldots\nonumber\\
&\succ&
\langle
0,\dimlev_{l-1},0,\ldots,0,
\dimlev_{l-1}+\dimlev_{l-1},
\dimlev_{l}+\dimlev_{l-1}\cdot\dimlev_{l}
\rangle\nonumber\\
&\equiv&
\langle
\underbrace{
0,\dimlev_{l-1},0,\ldots,0
}_{\dimlev_{l-1}},
2\cdot\dimlev_{l-1},
(1+\dimlev_{l-1})\cdot\dimlev_{l}
\rangle\nonumber\\
&&
\ldots
\text{after $j\cdot\dimlev_{l-1}$ steps of $\rewp$, from $\Pi$}
\ldots
\nonumber\\
&\succ&
\langle
\underbrace{
 \underbrace{
 0,\ldots,0
 }_{j},
 \dimlev_{l-1},0,\ldots,0
}_{\dimlev_{l-1}},
(j+1)\cdot\dimlev_{l-1},
(1+j\cdot\dimlev_{l-1})\cdot\dimlev_{l}
\rangle\nonumber\\
&&
\ldots
\text{after $\dimlev^2_{l-1}$ steps of $\rewp$, from $\Pi$}
\ldots
\nonumber
\\
%
&\succ&
\langle
\underbrace{0,\ldots,0}_{\dimlev_{l-1}},
(\dimlev_{l-1}+1)\cdot\dimlev_{l-1},
(1+\dimlev^2_{l-1})\cdot\dimlev_{l}
\rangle\nonumber\\
&&
\ldots
\text{after 
$(\dimlev_{l-1}+1)\cdot\dimlev_{l-1}+
(1+\dimlev^2_{l-1})\cdot\dimlev_{l}$ of $\rews\cup\rewl$-steps}
\ldots
\nonumber\\
&\succ&
\langle
0,\ldots,0
\rangle\nonumber\\
&\equiv&
\cutmeasure_l(\Pi')\nonumber\enspace ,
\end{eqnarray}
for some $\Pi'$. By all that means that we have just rewritten $\Pi$ to
$\Pi'$ after, at most,
$\dimlev^2_{l-1}+(\dimlev_{l-1}+1)\cdot\dimlev_{l-1}+
(1+\dimlev^2_{l-1})\cdot\dimlev_{l}\in
\dimlev_{l}(\Pi)\cdot\dimlev^2_{l-1}(\Pi)
\leq
2\cdot(
\maximaldimension(\Pi)+
\maximaldimension^2(\Pi)+
\maximaldimension^3(\Pi)
)
\leq
6\cdot\maximaldimension^3(\Pi)$ steps, since 
$\dimlev_{l}(\Pi)\leq\maximaldimension(\Pi)$, for every $0\leq l\leq
\maximaldepth(\Pi)$.

Finally, the third point.
If we find $\lblmeasure(\Pi')$, we get $\maximaldimension(\Pi')$ as well, which
is the sum of all the components of $\lblmeasure(\Pi')$.
Assume again to start from $\Pi$, and to rewrite it under $\strategy$.
We have:
\begin{eqnarray}
\lefteqn{\lblmeasure(\Pi)}\nonumber\\
&\equiv&
\langle
\dimlev_{0},
\ldots,
\dimlev_{l-2},
\dimlev_{l-1},
\dimlev_{l},
\ldots,
\dimlev_{\maximaldepth}
\rangle\nonumber\\
&&
\ldots\text{after $1$ step of $\rewp$, from $\Pi$}\ldots\nonumber\\
&\prec&
\langle
\dimlev_{0},
\ldots,
\dimlev_{l-2},
\dimlev_{l-1}+1,
\dimlev_{l}+\dimlev_{l},
\ldots,
\dimlev_{\mmm}+\dimlev_{\mmm}
\rangle\nonumber\\
&&
\ldots\text{after $i$ steps of $\rewp$, from $\Pi$}\ldots\nonumber\\
&\prec&
\langle
\dimlev_{0},
\ldots,
\dimlev_{l-2},
\dimlev_{l-1}+i,
\dimlev_{l}+i\cdot\dimlev_{l},
\ldots,
\dimlev_{\mmm}+i\cdot\dimlev_{\mmm}
\rangle\nonumber\\
&&
\ldots
\text{after $\dimlev_{l-1}$ steps of $\rewp$, from $\Pi$}
\ldots\nonumber\\
&\prec&
\langle
\dimlev_{0},
\ldots,
\dimlev_{l-2},
\dimlev_{l-1}+\dimlev_{l-1},
\dimlev_{l}+\dimlev_{l-1}\cdot\dimlev_{l},
\ldots,
\dimlev_{\mmm}+\dimlev_{l-1}\cdot\dimlev_{\mmm}
\rangle\nonumber\\
&\equiv&
\langle
\dimlev_{0},
\ldots,
\dimlev_{l-2},
2\cdot\dimlev_{l-1},
(1+\dimlev_{l-1})\cdot\dimlev_{l},
\ldots,
(1+\dimlev_{l-1})\cdot\dimlev_{\mmm},
\rangle\nonumber\\
&&
\ldots
\text{after $j\cdot\dimlev_{l-1}$ steps of $\rewp$, from $\Pi$}
\ldots
\nonumber\\
&\prec&
\langle
\dimlev_{0},
\ldots,
\dimlev_{l-2},
(j+1)\cdot\dimlev_{l-1},
(1+j\cdot\dimlev_{l-1})\cdot\dimlev_{l},
\ldots,
(1+j\cdot\dimlev_{l-1})\cdot\dimlev_{\mmm}
\rangle\nonumber
\end{eqnarray}

\begin{eqnarray}
&&
\ldots
\text{after $\dimlev^2_{l-1}$ steps of $\rewp$, from $\Pi$}
\ldots
\nonumber\\
&\prec&
\langle
\dimlev_{0},
\ldots,
\dimlev_{l-2},
(\dimlev_{l-1}+1)\cdot\dimlev_{l-1},
(1+\dimlev^2_{l-1})\cdot\dimlev_{l},
\ldots,
(1+\dimlev^2_{l-1})\cdot\dimlev_{\mmm}
\rangle\nonumber\\
&\equiv&
\lblmeasure(\Bar{\Pi})
\label{eqn:lbl-bound}
\enspace ,
\end{eqnarray}
for some $\Bar{\Pi}$.
At this point, $\Bar{\Pi}$ can be normalized at levels $l-1$, and
$l$ by reducing all $\rews\cup\rewl$-redexes which simply
erase structure.  We can safely state that, 
after (at most)
$(\dimlev_{l-1}+1)\cdot\dimlev_{l-1}+
(1+\dimlev^2_{l-1})\cdot\dimlev_{l}$
$\rews\cup\rewp$-steps, $\lblmeasure(\Bar{\Pi})$ here above
is a bound for $\lblmeasure(\Pi')$. It implies the third
point we want to prove.
\end{proof}

In a few we shall get the bound on the cut elimination complexity.
Thanks to Proposition~\ref{proposition:basic-complexity}
we can observe that each step $\strategy$ in:
\[
	 \Pi_0
         \strategy\ldots\strategy
         \Pi_i
	 \strategy
	 \Pi_{i+1}
         \strategy\ldots
\]
rewrites $\Pi_i$ in $\Pi_{i+1}$ using at most 
$6\cdot\maximaldimension^3(\Pi_i)$.
So, $\Pi_i$ is obtained after at most
\begin{eqnarray}
\label{eqn:maximaldepth}
\sum_{k=0}^{i} 
6^{\frac{3^k-1}{2}}\cdot
\maximaldimension^{3^k}(\Pi_0)
\end{eqnarray}
steps.
Fact~\ref{fact:maximaldepth} assures that the reduction sequence here above
can not be longer than $\maximaldepth(\Pi_0)$. In particular, it is shorter if
some $\rewrel_{ws}$-redexes erase, at some point, all the boxes
constituting the $\maximaldepth$-level of $\Pi_0$.
So, the upper limit of \eqref{eqn:maximaldepth} is $\maximaldepth(\Pi_0)$,
and we get:
\begin{eqnarray*}
\sum_{k=0}^{\maximaldepth(\Pi_0)} 
6^{\frac{3^k-1}{2}}\cdot
\maximaldimension^{3^k}(\Pi_0)
&\leq&
6^{\frac{3^{\maximaldepth(\Pi_0)}-1}{2}}\cdot
\sum_{k=0}^{3^{\maximaldepth(\Pi_0)}} \maximaldimension^{k}(\Pi_0)
\\
&\leq&
6^{\frac{3^{\maximaldepth(\Pi_0)}-1}{2}}\cdot
\frac{\maximaldimension^{3^{\maximaldepth(\Pi_0)}+1}(\Pi_0) - 1}
     {\maximaldimension(\Pi_0) - 1}
\\
&\in&
O(\maximaldimension^{3^{\maximaldepth}}(\Pi_0))
\enspace .
\end{eqnarray*}

\section{The Concrete Syntax}
\label{section:The Concrete Syntax}

Figure~\ref{figure:patterns}
\begin{figure}[htbp]
\begin{eqnarray*}
\mathsf{p} &::=& \TermVars\ \mid\ \mathsf{p}\tensor \mathsf{p}
\end{eqnarray*}
\caption{The patterns for the concrete syntax}
\label{figure:patterns}
\end{figure}
introduces the patterns of our concrete syntax.
The set of patterns is ranged over by $\mathsf{P}$, while
$\TermVars$ is ranged over by $x, y, w, z$.

Figure~\ref{figure:concrete-syntax}
\begin{figure}[htbp]
\begin{eqnarray*}
M, N &::=& \TermVars \mid\ 
(\lambda \mathsf{P}.M)\ \mid\ (MN)\ \mid\ M\tensor N\ \mid\ !M\ \mid\ \DB M\
\mid\ \S M\ \mid\ \DP M
\end{eqnarray*}
\caption{The concrete syntax}
\label{figure:concrete-syntax}
\end{figure}
defines the set $\Terms$ of the functional terms which we take as
concrete syntax.

For any pattern $x_1\tensor\ldots\tensor x_n$, the set
$\FV{x_1\tensor\ldots\tensor x_n}$ of its free
variables is $\{x_1,\ldots,x_n\}$.
As usual, $\lambda$ binds the variables of $M$ so
that $\FV{\lambda \mathsf{P}.M}$ is $\FV{M}\setminus\FV{\mathsf{P}}$.
The free variable sets of all the remaining terms are
obvious as the constructors $\otimes, !, \S, \DB$, and
$\DP$ do not bind variables. Both $!$ and $\S$ build
$!$-boxes and $\S$-boxes, respectively, being $M$ the
\emph{body}. The term constructor $\DB$ can mark one of the
\emph{entry points}, namely the inputs,
of both $!$-boxes, and $\S$-boxes, while $\DP$
can mark only those of  $\S$-boxes. 

We shall adopt the usual shortening for $\lambda$-terms:
$\lambda x_1. \ldots\lambda x_n.M$ is abbreviated by 
$\lambda x_1 \ldots x_n.M$, and $(M_1\ldots(M_n N)\ldots )$ by
$M_1\ldots M_n N$, \ie\ the application is left-associative by default.

The elements of $\Terms$ are considered up to
the usual $\alpha$-equivalence. It allows the
renaming of the bound variables of a term $M$. For example,
$!(\lambda x . (\DB y)\  x)$ and $!(\lambda z . (\DB y)\ z)$
are each other $\alpha$-equivalent.

The substitution of $M$ for
$x$ in $N$ is denoted by $N\sug{M}{x}$. It is the obvious extension
to $\Terms$ of the capture-free substitution of terms
for variables, defined for the $\lambda$-Calculus.
For example, $y\sug{x}{y}$ yields $y$. 

The substitutions can be generalized to $\msug{M_1}{x_1}{M_n}{x_n}$,
which means the simultaneous replacement of $M_i$ for $x_i$, for every
$1\leq i\leq n$.

We shall use $\equiv$ as syntactic coincidence.

\section{The Type Assignment}
\label{section:The Type Assignment}

We decorate the sequent calculus of Intuitionistic Light
Affine Logic with the terms of the concrete syntax.
So, the language of logical formulas and the sequent calculus we refer
to are those in Figure~\ref{figure:ILAL}.

Call \emph{basic set of assumptions} any set of pairs
$\{x_1: A_1,\ldots,x_n: A_n\}$ that can be seen as a function
with finite domain $\{x_1,\ldots,x_n\}$. Namely, if $i\neq j$,
then $ x_i\neq x_j$.

An \emph{extended set of assumptions} is a basic set,
containing also pairs $\mathsf{P}: A$, that satisfies some further
constraints. A pattern
$\mathsf{P}\equiv
x_{1}\!\otimes\!\ldots\!\otimes\! x_{m}:A$ belongs to an extended set
of assumptions:
\begin{enumerate}
\item
if $A$ is $ A_{1}\!\otimes\!\ldots\!\otimes\! A_{p}$,
with $p\geq m$, and
\item
if $\{x_{1}: B_1,\ldots,x_{m}: B_m\}$ is a basic set of assumptions,
where every $B_i$ is either a single formula, or  tensor of formulas.
\end{enumerate}
For example, $\{x:\C,y:\B\}$ is a legal extended set, while
$\{z\!\otimes\! x:\C,y:\B\}$ is not.

Talking about ``assumptions'', we generally mean
``extended set of assumptions''.
Meta-variables for ranging over the assumptions are $\Gamma$,
 and $\Delta$.

The \emph{substitutions} on formulas replace formulas 
for variables in the obvious way.

Figure~\ref{figure:ILAL-decorated}
\begin{figure}[hbtp]
\begin{eqnarray*}
  \dedrule{(Ax)}{}{ x: B\TS  x: B}
\qquad
  \dedrule{(Cut)}
        {\Gamma\TS M: A
         \quad
         \Delta, x: A\TS N: B}
        {\Gamma,\Delta\TS N\sug{M}{ x}: B}
\end{eqnarray*}

\begin{eqnarray*}
  \dedrule{(Weak.)}
        {\Gamma \TS M: B}
        {\Gamma, x: A\TS M: B}
\qquad
  \dedrule{(Contr.)}
        {\Gamma,x:! A,y:! A\TS M: B}
        {\Gamma,z:! A\TS M\{{}^{z}\!\downarrow\!{}_{x}\
                                {}^{z}\!\downarrow\!{}_{y}\}: B}
\end{eqnarray*}

\begin{eqnarray*}
  \dedrule{(\linimpl_l)}
  {\Gamma \TS M:  A 
   \qquad
   \Delta, y: B \TS N: C}
  {\Gamma, \Delta, x: A\linimpl B \TS N\sug{xM}{ y}: C}
&\quad&
  \dedrule{(\linimpl_r)}
  {\Gamma, \mathsf{P}: B_1\!\otimes\!\ldots\!\otimes\! B_n \TS M: B}
  {\Gamma \TS 
   \lambda \mathsf{P}. M:
    B_1\!\otimes\!\ldots\!\otimes\! B_n \linimpl  B}
\end{eqnarray*}

\begin{eqnarray*}
  \dedrule{(\otimes_l)}
  {\Gamma,  x_1: B_1,  x_2: B_2 \TS M: B}
  {\Gamma,  x_1\!\otimes\! x_2: B_1\!\otimes\! B_2 \TS M: B}
&\quad&
  \dedrule{(\otimes_r)}
  {\Gamma \TS M: B \quad \Delta \TS N: A}
  {\Gamma, \Delta \TS M\!\otimes\! N: B\!\otimes\! A}
\end{eqnarray*}

\begin{eqnarray*}
  \dedrule{(!)}
  {\ldots  x_i: A_i
   \ldots \TS M: B
   \qquad 0\leq i \leq n \leq 1}
  {\ldots x_i:! A_i
   \ldots \TS !M
         \{\cdots{}^{\DB x_i}\!\!\downarrow\! {}_{ x_i }\cdots\}
   :! B}
\end{eqnarray*}

\begin{eqnarray*}
  \dedrule{(\S)}
  {\ldots  x_i : B_i
   \ldots  x'_j: A_j
   \ldots \TS M: B
   \qquad 0\leq i \leq m \quad 0\leq j\leq n
   }
  {\ldots x_i:!  B_i
   \ldots x'_j:\S A_j\ldots
   \TS
   \S M \{\cdots{}^{\DB x_i}\!\!\downarrow\! {}_{ x_i }
          \cdots{}^{\DP x'_j}\!\!\downarrow\! {}_{ x'_j}\cdots\}
   :\S B
  }
\end{eqnarray*}

\begin{eqnarray*}
  \dedrule{(\forall_l)}
        {\Gamma,  x:\sug{ B}{\VS} A \TS M: B}
        {\Gamma,  x:\forall \VS. A \TS M: B}
&\qquad&
 \dedrule{(\forall_r)}
        {\Gamma \TS M: A\qquad
         \VS\not\in\FV{\Gamma} }
        {\Gamma \TS M:\forall \VS. A}
\end{eqnarray*}
\caption{Decorating the sequent calculus with terms}
\label{figure:ILAL-decorated}
\end{figure}
introduces the sequent calculus of ILAL, decorated
with the terms of $\Terms$. 
Observe that $(!)$-rule can have at most one assumption.
Observe also that the two rules for the second order formulas
are not encoded by any term. Namely, we introduce a system
analogous to Mitchell's language \emph{Pure Typing Theory}
\cite{Mitchell:InfComp88}.
In this case, the logical system of reference is second order
ILAL, in place of System $\mathcal{F}$ \cite{GLT:PT}.

\section{The Dynamics for the Concrete Syntax}
\label{section:The Dynamics Concrete Syntax}

Figure~\ref{figure:rewriting-system}
\begin{figure}[htbp]
\begin{eqnarray*}
(\lambda x_1\!\otimes\!\ldots\!\otimes\! x_{m\geq 1}.M)
M_1\!\otimes\!\ldots\!\otimes\! M_m\ 
&\rhd_{\beta}& 
M\msug{M_1}{x_1}{M_m}{x_m}
\\
\DB !M   &\rhd_{!}& M
\\
\DP \S M &\rhd_{\S}& M
\end{eqnarray*}
\caption{The rewriting relations for the concrete syntax}
\label{figure:rewriting-system}
\end{figure}
\noindent
defines the basic rewriting relations on $\Terms$.

The first relation is the trivial generalization of the $\beta$-rule of
$\lambda$-Calculus to abstractions that bind patterns which represent
tuples of variables.
The $\alpha$-equivalence must be used to avoid variable
clashes when rewriting terms.
The second rewriting relation \emph{merges} the borders of two boxes.

Define the rewriting system $\rewsyst$ as the contextual
closure on $\Terms$ of the rewriting relations in
Figure~\ref{figure:rewriting-system}.
Its reflexive, and transitive closure is $\rewsyst^*$.
The pair $(\Terms,\rewsyst)$ is the functional language we
shall use to prove P-Time completeness of ILAL. We shall generally abuse
the notation by referring to such a language only with $\Terms$.

\section{Comments on the Concrete Syntax}
\label{section:Comments on the Concrete Syntax}

$\Terms$ gives a very compact representation of the derivations.
The contraction is represented by multiple occurrences of the
same variable. The pattern matching avoids the use of
any $let$-like binder that would require to extend $\rewsyst$ by some
commuting conversions.
The boxes have not any interface like in the paradigmatic language
proposals of \cite{Asperti:LICS98,Roversi:ASIAN98,Roversi:IJFCS00}.

\begin{figure}
\centering\epsfig{file=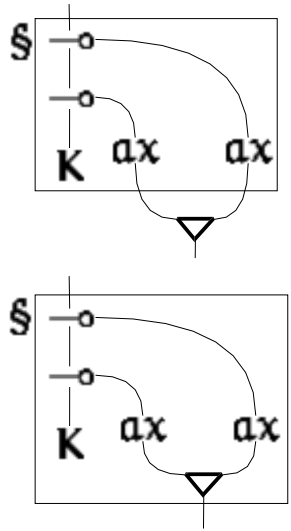,
bbllx=0, bblly=386, bburx=115, bbury=596, clip=}
\caption{Two nets for the same term}
\label{figure:two-repr-net}
\end{figure}

However, we have to pay for this notational economy.
The typable sub-set of $\Terms$ is not at all an isomorphic representation of
the derivations. 
The simplest example to observe how ambiguously $\Terms$ represents ILAL
is in Figures~\ref{figure:two-repr-der},
\begin{figure}
\begin{eqnarray*}
\dedrule{(Contr.)}
{\dedrule{(\S)}
 {x_1:!\alpha, x_2:!\alpha \vdash K\ x_1\ x_2:!\alpha}
 {w_1:!!\alpha, w_2:!!\alpha \vdash \S(K\ \DP w_1\ \DP w_2):\S !\alpha}
}
{z:!!\alpha \vdash \S(K\ \DP z\ \DP z):\S !\alpha}
\end{eqnarray*}
\begin{eqnarray*}
\dedrule{(\S)}
{\dedrule{(Contr.)}
 {x_1:!\alpha, x_2:!\alpha \vdash K\ x_1\ x_2:!\alpha}
 {w:!\alpha \vdash K\ w\ w:!\alpha}}
{z:!!\alpha \vdash \S(K\ \DP z\ \DP z):\S !\alpha}
\end{eqnarray*}
\caption{Two derivations for the same term}
\label{figure:two-repr-der}
\end{figure}
 and~\ref{figure:two-repr-net}.
The same term $\S(K\ \DP z\ \DP z)$ ``encodes'' two radically different
derivations of the sequent calculus, \ie\
$\S(K\ \DP z\ \DP z)$ ``encodes'', under the same order as in
Figure~\ref{figure:two-repr-der}, the two nets in
Figure~\ref{figure:two-repr-net}.

Here we want to stress that such an ambiguity is \emph{not} an issue for us.
The concrete syntax is not meant to be a real calculus,
but just a compact notation for proofs.
What we need is a language where we can observe the type discipline at work,
especially in the proof of P-Time completeness.
In case we want to evaluate $M\in\Terms$ with polynomial cost,
the right way to do it is to translate $M$ into a proof net, so that 
$\strategy$ and the good computational properties of the nets can be exploited.

We only need to agree about the translation from $\Terms$ to the
nets. We choose the one putting the contractions as deeply as possible.
So we would adopt the lower most net in Figure~\ref{figure:two-repr-net}
as a translation of $\S(K\ \DP z\ \DP z)$.
This choice reduces the computational complexity of the translation.

\section{Encoding a Numerical System}
\label{section:Numerical System}

The numerical system adopted on $\Terms$ is the analogous
of Church numerals for $\lambda$-Calculus.

The type and the terms of the \emph{tally} integers are in
Figure~\ref{figure:tally-integers}.
\begin{figure}[htbp]
\begin{eqnarray*}
\Int&=&
\forall\alpha. 
!(\alpha \linear \alpha)\linear\S(\alpha \linear \alpha)
\\
\Zero&=&\lambda x.\S\lambda y.y:\Int\\
\Num{n}&=&
\lambda x.\S(\lambda y.\underbrace{\DB x(\ldots(\DB x}_n\ y)\ldots)):\Int
\enspace .
\end{eqnarray*}
\caption{The tally integers}
\label{figure:tally-integers}
\end{figure}
Observe that there is a translation from $\Terms$ to $\lambda$-Calculus
that, applied to $\Zero$ and $\Num{n}$, yields 
$\lambda$-Calculus Church numerals:
\begin{eqnarray*}
&&\lambda fx.\underbrace{f(\ldots(f}_{n\geq 0}\ x)\ldots))
\enspace .
\end{eqnarray*}
The translation just erases all the occurrences of
$!, \S, \DB$, and $\DP$.

Figure~\ref{figure:combinators-on-integers}
\begin{figure}[htbp]
\begin{eqnarray*}
\Succ&=&\lambda zx.\S(\lambda y.\DB x(\DP(z\ x)\ y)):
\Int\linimpl\Int\\
\Sum&=&\lambda wzx.\S(\lambda y.\DP(w\ x)(\DP(z\ x)\ y)):
\Int\linimpl\Int\linimpl\Int\\
\Iter&=&\lambda x y z.\S(\DP(x\ y)\ \DP z):
\Int\linimpl!(A\linimpl A)\linimpl\S A\linimpl\S A\\
\Mult&=&\lambda xy.
           \Iter\ x\
           !(\lambda w.\Sum\ \DB y\ w)\
           \S \Zero:
\Int\linimpl!\Int\linimpl\S\Int\\
\Coerc&=&\lambda x.\S(\DP(x\ !\Succ)\ \Zero):
\Int\linimpl\S\Int
\enspace .
\end{eqnarray*}
\caption{Some combinators on the tally integers}
\label{figure:combinators-on-integers}
\end{figure}
introduces some further combinators on the numerals.
The numeral next to $\Num{n}$ can be calculated as in
Figure~\ref{figure:next-integer}.
\begin{figure}[htbp]
\begin{eqnarray*}
\lefteqn{\Succ\ \Num{n}}\\
&\rewsyst&
\lambda x.\S(\lambda y.\DB x(\DP(\Num{n}\ x)\ y))\\
&\rewsyst&
\lambda x.
\S(\lambda y.\DB x(
                   \DP\S(\lambda w.\overbrace{\DB x(\ldots(\DB x}^n\ w)\ldots))
                   \ y
                  )
  )\\
&\rewsyst&
\lambda x.
\S(\lambda y.\DB x(
                   (\lambda w.\DB x(\ldots(\DB x\ w)\ldots))
                   \ y
		  )
  )\\
&\rewsyst&
\lambda x.
\S(\lambda y.\underbrace{\DB x(\DB x(\ldots(\DB x}_{n+1}\ y)\ldots))
  )\\
&=& \Num{n+1}\enspace .
\end{eqnarray*}
\caption{Calculating the numeral next to $\Num{n}$}
\label{figure:next-integer}
\end{figure}
$\Sum$ adds two numerals.
$\Iter$ takes as arguments a numeral, a \emph{step} function,
and a \emph{base} where to start the iteration from.
Observe that $\Iter\ \Num{2}\ !\Num{n}\ \S\Zero$ cannot have type,
for any numeral $\Num{n}$.
This because the step function is required
to have identical domain and co-domain. 
This should not surprise. Taking the $\lambda$-Calculus Church numeral
$\Num{2}$, and applying it to itself we get an exponentially costing
computation.

$\Mult$ is defined as an iterated sum, for multiplying two numerals.

Finally, $\Coerc(\text{ion})$ 
embeds a numeral into a $\S$-box, preserving its value. Look at
Figure~\ref{figure:coercion-compuatation} for an example.
\begin{figure}[htbp]
\begin{eqnarray*}
\lefteqn{\Coerc\ \Num{n}}\\
&\rewsyst&
\S(\DP(\Num{n}\ !\Succ)\ \Zero)\\
&\rewsyst&
\S(\DP(\S(\lambda w.\DB !\Succ(\ldots(\DB !\Succ\ w)\ldots))
      )\ \Zero)\\
&\rewsyst&
\S((\lambda w.\Succ(\ldots(\Succ\ w)\ldots))
   )\ \Zero)\\
&\rewsyst&
\S(\overbrace{\Succ(\ldots(\Succ}^n\ \Zero)\ldots))\\
&\rewsyst^*&
\S\Num{n}
\enspace .
\end{eqnarray*}
\caption{Coercion of $\Num{n}$ to $\S\Num{n}$}
\label{figure:coercion-compuatation}
\end{figure}

       \subsection{Encoding a Predecessor}
\label{subsection:The Predecessor}

The predecessor of the numerical system for $\Terms$ is an instance of a
general computation scheme that iterates the template function in
Figure~\ref{figure:template-function}.
\begin{figure}[htbp]
\begin{eqnarray*}
\mathcal{T}_f(g,h)&=& (f, gh)
\enspace .
\end{eqnarray*}
\caption{Template function for the predecessor}
\label{figure:template-function}
\end{figure}
$\mathcal{T}$ takes a pair of functions $h,g$ as arguments, and has $f$
as its parameter.
If $h:X\rightarrow Y, g:Y\rightarrow Z$, and $f:Z\rightarrow Z$,
for some domains $X, Y, Z$,
then $\mathcal{T}_f$ can be iterated. An example of an $n$-fold iteration 
of $\mathcal{T}_f$ from $(g,h)$ is in 
Figure~\ref{figure:n-long-iteration},
\begin{figure}[htbp]
\begin{eqnarray*}
\overbrace{\mathcal{T}_f(\ldots\mathcal{T}_f}^n(gh)\ldots) &=&
(f,\overbrace{f(\ldots f}^{n-1}(gh)\ldots))
\end{eqnarray*}
\caption{Iterating the template function}
\label{figure:n-long-iteration}
\end{figure}
where it is simple to recognize the predecessor of $n$, if we let
$f:\mathbb{N}\rightarrow\mathbb{N}$ be the identity,
$g:\mathbb{N}\rightarrow\mathbb{N}$ be the successor, $h$ be $0$, and if
we assume to erase the first component of the result.
Recasting everything in $\Terms$, we get the definitions in
Figure~\ref{figure:predecessor}. 

\begin{figure}[htbp]
\begin{eqnarray*}
I&=&\lambda x.x:\forall\alpha.\alpha\linimpl\alpha\\
\pi_2&=&\lambda x\otimes y.y:
  \forall\alpha.\alpha\otimes\alpha\linimpl\alpha\\
T&=&\lambda f.\lambda g\otimes h. (f\otimes (gh)):
\forall \alpha.
(\alpha\linimpl\alpha)
\linimpl
((\alpha\linimpl\alpha)\otimes\alpha)
\linimpl
(\alpha\linimpl\alpha)
\otimes
\alpha\\
\Step\ z&=&T\ z:
((\Int\linimpl\Int)\otimes\Int)
\linimpl
(\Int\linimpl\Int)
\otimes\Int\\
\Base\ y&=&T\ I\ (I\otimes y):
(\Int\linimpl\Int)\otimes\Int
\\
\Pred&=&
\lambda wx.
\S(\lambda y.\pi_2(\DP(w\ !(\Step\ \DB x))(\Base\ y)))
:\Int\linimpl\Int
\\
\text{where }&& y, z:\Int
\end{eqnarray*}
\caption{The predecessor}
\label{figure:predecessor}
\end{figure}
The term $\Pred$ iterates $w$ times
$!(\Step\ \DB x)$ from $I\otimes y$, exploiting the correspondence
between $\mathcal{T}$ and $T$ in Figure~\ref{figure:correspondence}.
\begin{figure}[htbp]
\begin{tabular}{|c|c|}
\hline
$T$ & $\mathcal{T}$\\ \hline\hline
$!(\Step\ \DB x)$ & $f$\\ \hline
$I$              & $g$\\ \hline
$I\tensor y$     & $h$\\
\hline
\end{tabular}
\caption{Correspondence between $\mathcal{T}$ and $T$}
\label{figure:correspondence}
\end{figure}
Observe also that our predecessor does not make any explicit use of the
encoding of the additive types by means of the second order quantification.
In \cite{Asperti:LICS98} the predecessor has a somewhat more intricate
form that we recall here:
\begin{eqnarray}
\label{eqn:old-pred}
\lambda nxy.(n\ (\lambda p.(U\ I\ x\ (p\ \Snd)))\ (U\ I\ I\ y)\ \Fst)
\enspace ,
\end{eqnarray}
where:
\begin{eqnarray*}
U\ P\ Q\ R &=& \lambda z.(z\ P\ Q\ R)\\
\Fst &=& \lambda x y z. (x\ z)\\
\Snd &=& \lambda x y z. (y\ z)
\enspace .
\end{eqnarray*}
$\Pred$ is obtained by eliminating the non essential
components of \eqref{eqn:old-pred} here above.

Both $\Pred$, and \eqref{eqn:old-pred} are 
\emph{syntactically linear}, so also their complexity is readily linear.
On the contrary, the usual encoding of the predecessor, that, using
$\lambda$-Calculus syntax with pairs $\langle M, N \rangle$, is:
\begin{eqnarray}
\label{eqn:classic-pred}
\lambda nxy.
\Fst(n\ (\lambda p.\langle\Snd(p),x\ \Snd(p)\rangle)\ \langle y,y\rangle)
\enspace ,
\end{eqnarray}
has also an exponential strategy.
Such a strategy exists because the term is not syntactically linear.
However both $\Pred$, and \eqref{eqn:old-pred} witness that the non
linearity of \eqref{eqn:classic-pred} is inessential. In particular, 
in \cite{Girard:LLL98}, where Girard embeds \eqref{eqn:classic-pred}
in LLL, the sub-term $\lambda p.\langle\Snd(p),x\ \Snd(p)\rangle$
here above has the \emph{additive} type
$(\alpha\&\alpha) \linear (\alpha\&\alpha)$.
This means that, at every step of the iteration
$n\ (\lambda p.\langle\Snd(p),x\ \Snd(p)\rangle)\ \langle y,y\rangle$,
only one of the multiple uses of $p$ is effectively
useful to produce the result. 

\begin{remark}
\begin{itemize}
\item
The procedural iteration scheme in Figure~\ref{figure:n-long-iteration}, our
predecessor is an instance of, was already used in \cite{Roversi:ASIAN98}.
However, only reading \cite{DanosJoinet:ICC99}, we saw that the iteration
in Figure~\ref{figure:n-long-iteration} actually ``implements''
a general logical iteration scheme, which we adapt to ILAL in
Figure~\ref{figure:gen-iter-scheme}. There,
the term $M$ must contain $g\ h$, the argument of $n\ \mathcal{T}_f$.
\begin{figure}[htbp]
\begin{eqnarray*}
\dedrule{(\linear_l)}
{
!\Gamma \vdash \mathcal{T}_f:!(A\linear A)
\qquad\qquad\qquad
\Delta, y:\S(A\linear A)\vdash M:B
}
{
 \dedrule{(\forall_l)}
 {
 !\Gamma,\Delta,
 n:(!(\alpha\linear \alpha)\linear\S(\alpha\linear \alpha))\sug{A}{\alpha}
 \vdash M\sug{\DP(n\ \mathcal{T}_f)}{y}:B
 }
 {
 !\Gamma,\Delta, \Int \vdash M\sug{\DP(n\ \mathcal{T}_f)}{y}:B
 }
}
\end{eqnarray*}
\caption{General iteration scheme: the logical structure}
\label{figure:gen-iter-scheme}
\end{figure}
The more traditional iteration scheme can be obtained from 
Figure~\ref{figure:gen-iter-scheme} by letting 
$!\Delta, y:\S(A\linear A)\vdash M:B$ be the conclusion of the derivation in
Figure~\ref{figure:inst-gen-iter-scheme}.
\begin{figure}[htbp]
\begin{eqnarray*}
\dedrule{(\linear_l)}
{
\Theta \vdash N:A
\qquad\qquad
x:A \vdash x:A
}
{
 \dedrule{(\S)}
 {
 \Theta, y:A\linear A\vdash y\ N:A
 }
 {
 \S\Theta, w:\S(A\linear A) \vdash \S(\DP w\ N):\S A
 }
}
\end{eqnarray*}
\caption{Getting the standard iteration scheme}
\label{figure:inst-gen-iter-scheme}
\end{figure}
Observe that the instance of $M$ we use for our predecessor is not as simple as
$\S(\DP w\ N)$.
\item
We want to discuss a little more about the linearity of the additive structures.
Not sticking to any particular notation, let
$fxy = \Fst\langle xy,xy\rangle$. The function $f$ is just the identity,
and it would get a linear type in ILAL. Now, consider an $n$-fold iteration
of $f$ by means of a Church numeral $\Num{n}$ .
Then, let us apply the result to a pair of identities.
We have just defined $g_f\ n = ((n\ f)\ I)\ I$. This term is typable
in ILAL. So, in ILAL,  $g_f\ n$ normalizes in polynomial, actually linear, time. 
However, try to reduce $g_f\ n$ in most
traditional lazy call-by-value implementations of functional languages
(SML, CAML, Scheme, etc.), you will discover that the
reduction takes exponential time. So,
firstly, if a usual $\lambda$-term $M$ can be embedded 
in ILAL, then, in general, it is \emph{not} true that $M$ normalizes in
polynomial time under \emph{any} reduction strategy.
We only know that \emph{there exists} an effective
way to normalize $M$ in polynomial time. The polynomial reduction, in general,
is  \emph{not compatible} with the \emph{lazy call-by-value reduction}.

However, consider again $g_f\ n = ((n\ f)\ I)\ I$ and evaluate it
under the lazy call-by-name strategy: it will cost linear time.
We leave the following open question: is it true that, taking a typable term 
$M$ having a polynomial reduction strategy, then that strategy can
be the lazy call-by-name?
\end{itemize}
\end{remark}

\section{Encoding the Polynomials}
\label{section:Polynomials}

In this section we show how to encode the elements of $\mathcal{P}$, \ie\
the polynomials with positive degrees, and positive coefficients, as terms
of $\Terms$. This encoding is based on the numerical system of
Section~\ref{section:Numerical System}. It will serve to represent and simulate
all P-Time Turing machines using the terms of ILAL.

We use $p^\vartheta_x$ to range over the polynomials
$\sum_{i=0}^\vartheta a_i x^i\in\mathcal{P}$
with maximal non null degree $\vartheta$, and indeterminate $x$.

The result of this section is:
\begin{theorem}
\label{theorem:polynomial-representation}
There is a translation 
$\Hat{\phantom{p}}:\mathcal{P}\rightarrow\Terms$, such that,
for any $p^\vartheta_x\in\mathcal{P}$: 
\begin{itemize}
\item 
$\Hat{p}^\vartheta_x:\Int\linimpl\S^{\vartheta+3}\Int$, and
\item
$p^\vartheta_n = m$,
if, and only if, 
$\Hat{p}^\vartheta_{\Num{n}}\rewsyst^*\S^{\vartheta+3}\Num{m}$.
\end{itemize}
\end{theorem}

In the following we develop the proof of the theorem, and an example about how
the encoding works.

First of all, some useful notations.

Let $p^{\vartheta}_x$ be the polynomial $\sum_{i=0}^\vartheta a_i x^i$ describing the
computational bound of the Turing machine being encoded.
Let $\kappa = \frac{\vartheta(\vartheta+1)}{2}$.

Abbreviate with $\vec{y}_n$ an $n$-long vector of all vectors with length $1$
through $n$, each containing variables $y^j_i\in\TermVars$, where
$0\leq i\leq n-1$, and $1\leq j\leq n$. 
Figure~\ref{figure:vector-of-vectors}
\begin{figure}[htbp]
\begin{eqnarray*}
\vec{y}_3 &=&
\begin{array}[t]{ccc}
y^1_0 & & \\
y^2_0 & y^2_1 & \\
y^3_0 & y^3_1 & y^3_2
\end{array}
\end{eqnarray*}
\caption{Vector of vectors of variables}
\label{figure:vector-of-vectors}
\end{figure}
gives $\vec{y}_3$ as an example.
As usual, $\vec{y}_3[i][j]$ picks $y^j_i$ out of the vector $\vec{y}$. 

Figure~\ref{figure:generalizing-integere-operations},
\begin{figure}[htpb]
\begin{eqnarray*}
\Int_n&=&\Int\otimes\ldots\otimes\Int
\qquad \text{with }n\text{ components}
\\
\Num{0}_n&=&\Num{0}\otimes\ldots\otimes\Num{0}
\qquad\qquad \text{with }n\text{ components}
\\
\lozenge^n M&=&
\underbrace{\lozenge(\ldots(\lozenge}_n M)\ldots)\qquad
 \text{with } \lozenge\in\{!,\S,\DB,\DP\}
\\
\Num{0}^{p,q}&=&\S^p !^q\Num{0}:\S^p !^q\Int\\\Sum_n&=&
\lambda x_1\tensor\ldots \tensor x_n z.
\S(\lambda y.
   \DP(x_1\ z)(\ldots(\DP(x_n\ z)\ y)\ldots)
):\Int_n\linimpl\Int
\\
\Sum^p_n&=&
\lambda x_1\tensor\ldots\tensor x_n .
\S^p(\Sum_n\ \DP^p x_1\tensor\ldots\tensor\DP^p x_n):(\S^p\Int)_n\linimpl\S^p\Int\\
\Succ^{p,q}&=&
\lambda x. \S^p(!^q(\Succ\ \DB^q(\DP^p x))):\S^p !^q\Int\linimpl\S^p !^q\Int\\
\Coerc^{p,q}&=&\lambda x.\S(\DP(x\ !\Succ^{p,q})\ \Num{0}^{p,q}):
\Int\linimpl\S^{p+1} !^q\Int\\
\Mult^p&=&
\lambda x y.\S^p(\Mult\ \DP^p x\ \DP^p y ):
\S^p\Int\linimpl\S^p !\Int\linimpl\S^{p+1}\Int\\
\Tuple_n&=&
\lambda x.\S(\DP(x\ !(\lambda x_1\tensor\ldots\tensor x_n.
                      \Succ\ x_1\tensor\ldots\tensor\Succ\ x_n))\ \Num{0}_n):\\
&&\qquad\qquad\qquad\qquad\qquad\qquad\qquad\qquad
\qquad\qquad\qquad\quad
\Int\linimpl\S(\Int_n)
\end{eqnarray*}
\caption{Generalizations of operations on the numerals}
\label{figure:generalizing-integere-operations}
\end{figure}
where $p,q\geq 0$, and $n\geq 1$, introduces both a type abbreviation,
and some generalizations of the operations on Church numerals in
Section~\ref{section:Numerical System}.

Figure~\ref{figure:polynomial-encoding}
\begin{figure}[htpb]
\begin{eqnarray*}
\Hat{p}^\vartheta_x
&=&
\lambda x.
\S((\lambda \begin{array}[t]{l}
            y^1_0\\
            \tensor\ldots\tensor\\
            y^i_0\;\tensor\ldots\tensor y^i_{i-1}\\
	    \tensor\ldots\tensor\\
	    y^\vartheta_0\tensor\ldots\tensor y^\vartheta_{\vartheta-1}.
	    \end{array}\\
&&\phantom{\lambda x.\S(( }
\Sum^{\vartheta+2}_{\vartheta+1}\
\S^{1}(\Coerc^{\vartheta,0}\  \DP^{1}
     \doublelangle a_0 x^0\doublerangle_{\vec{y}_{\vartheta}[0]})\\
&&\phantom{\lambda x.\S(( \Sum^{\vartheta+1}_{\vartheta+1}\ }
\qquad\qquad\vdots\\
&&\phantom{\lambda x.\S(( \Sum^{\vartheta+1}_{\vartheta+1}\ }
\tensor
\S^{i+1}(\Coerc^{\vartheta-i,0}\ \DP^{i+1}
     \doublelangle a_i x^i\doublerangle_{\vec{y}_{\vartheta}[i]})\\
&&\phantom{\lambda x.\S(( \Sum^{\vartheta+1}_{\vartheta+1}\ }
\qquad\qquad\vdots\\
&&\phantom{\lambda x.\S(( \Sum^{\vartheta+1}_{\vartheta+1}\ }
\tensor
\S^{\vartheta+1}(\Coerc^{0,0}\ \DP^{\vartheta+1}
     \doublelangle a_\vartheta 
                   x^\vartheta\doublerangle_{\vec{y}_{\vartheta}[\vartheta]})\\
&&\phantom{\lambda x.\S( }
   )\ \DP(\Tuple_{\kappa}\ x)):\Int\linimpl\S^{\vartheta+3}\Int
\\
\\
\text{where:}
\\
\doublelangle a x^0\doublerangle_{\vec{z}}
     &\mapsto&\Coerc^{0,0}\ \Num{a}:\S\Int\\
\doublelangle a x^n\doublerangle_{\vec{z}}
     &\mapsto&\Mult^n\ \langle \vec{z},n-1\rangle\
       (\Coerc^{n-1,1}\ \Num{a}):\S^{n+1}\Int
\qquad(n\geq 1)
\\
\\
\langle \vec{z},0\rangle&\mapsto&\Coerc^{0,0}\ \vec{z}[0]:\S\Int\\
\langle \vec{z},n\rangle&\mapsto&
\Mult^{n}\ \langle\vec{z},n-1\rangle\ (\Coerc^{n-1,1}\ \vec{z}[n]):
\S^{n+1}\Int
\qquad(n\geq 1)
\end{eqnarray*}
\caption{Encoding of the polynomial}
\label{figure:polynomial-encoding}
\end{figure}
encodes the polynomial $p^\vartheta_x$, on which we 
can remark some simple facts.
$\Tuple_n$ makes $n$ copies of the numeral it is applied to.
Every ``macro''
$\doublelangle a_i\cdot x^i\doublerangle_{\vec{y}_{\vartheta}[i]}$ represents
the factor $a_i x^i$ so that $x^i$ is a product of as many variables of
$\TermVars$ as the degree $i$. The coercion applied to each of them just adds as many
$\S$-boxes as necessary to have all the arguments of
$\Sum^{\vartheta+2}_{\vartheta+1}$ at the same depth $\vartheta+2$.

We conclude this section with an example.
Figure~\ref{figure:full-polynomial-encding}
\begin{figure}[htpb]
\begin{eqnarray*}
\Hat{p}^{2}_{\Num{2}}
&=&
\S((\lambda y^1_0\tensor y^2_0\tensor y^2_1.\\
&&\phantom{\S(( }
\Sum^{4}_{3}\
\S^{1}(\Coerc^{2,0}\  \DP^{1}
     \doublelangle 1\cdot x^0\doublerangle_{\vec{y}_2[0]})\\
&&\phantom{\S(( \Sum^{4}_{3}\ }
\qquad\qquad\tensor\\
&&\phantom{\S(( \Sum^{4}_{3}\ }
\S^{2}(\Coerc^{1,0}\ \DP^{2}
     \doublelangle 0\cdot x^1\doublerangle_{\vec{y}_2[1]})\\
&&\phantom{\S(( \Sum^{4}_{3}\ }
\qquad\qquad\tensor\\
&&\phantom{\S(( \Sum^{4}_{3}\ }
\S^{3}(\Coerc^{0,0}\ \DP^{3}
     \doublelangle 1\cdot x^2\doublerangle_{\vec{y}_2[2]})\\
&&\phantom{\S( }
   )\ \DP(\Tuple_{3}\ \Num{2})):\Int\linimpl\S^{5}\Int
\\
\\
\text{where:}&&
\\
\doublelangle 1\cdot x^0\doublerangle_{\vec{y}_2[0]}
     &=&\Coerc^{0,0}\ \Num{1}:\S\Int
     \\
\doublelangle 0\cdot x^1\doublerangle_{\vec{y}_2[1]}
     &=&\Mult^1\ ( \Coerc^{0,0}\ \vec{y}_2[0][1]  )\
        (\Coerc^{0,1}\ \Num{0}):\S^{2}\Int
     \\
\doublelangle 1\cdot x^2\doublerangle_{\vec{y}_2[2]}
     &=&\Mult^2\ 
        ( \Mult^1\ ( \Coerc^{0,0}\ \vec{y}_2[0][2])\\
&&\phantom{\Mult^2\ ( \Mult^1\ }
                   (\Coerc^{0,1}\ \vec{y}_2[1][2])\\
&&\phantom{\Mult^2\ }
        )\ (\Coerc^{1,1}\ \Num{1}):\S^{3}\Int\enspace ,
\\
\text{and}&&
\\
\vec{y}_2[0][1]&=&y^1_0\\	
\vec{y}_2[0][2]&=&y^2_0\\	
\vec{y}_2[1][2]&=&y^2_1
\end{eqnarray*}
\caption{Encoding the  polynomial $x^2+1$}
\label{figure:full-polynomial-encding}
\end{figure}
fully develops the encoding of the polynomial $x^2+1$.
Assume we want to evaluate $\Hat{p}^{2}_{\Num{2}}$,
from which we expect $\S^5\Num{5}$. Figure~\ref{figure:intermediate-steps}
\begin{figure}[htpb]
\begin{eqnarray*}
\DP(\Tuple_{3}\ \Num{2}) &=&\Num{2}\tensor\Num{2}\tensor\Num{2}
\\
\DP^1(\Coerc^{0,0}\ \Num{1})&=&\Num{1}
\\
\DP^2(\Mult^1\ ( \Coerc^{0,0}\ \Num{2}  )\
         (\Coerc^{0,1}\ \Num{0}))&=&\Num{0}
\\
\DP^3(\Mult^2\ ( \Mult^1\ ( \Coerc^{0,0}\ \Num{2}) &&
\\
\phantom{\Mult^2\ ( \Mult^1\ }
                    (\Coerc^{0,1}\ \Num{2})&&
\\
\phantom{\Mult^2\ }
        )\ (\Coerc^{1,1}\ \Num{1}))&=&\Num{4}
\\
\S^{1}(\Coerc^{2,0}\  \Num{1})&=&\S^{4}\Num{1}\\
\S^{2}(\Coerc^{1,0}\ \Num{0})&=&\S^{4}\Num{0}\\
\S^{3}(\Coerc^{0,0}\ \Num{4})&=&\S^{4}\Num{4}\\
\S(\Sum^{4}_{3}\ (\S^{4}\Num{1}\otimes\S^{4}\Num{0}\otimes\S^{4}\Num{4}))&=&
\S^{5}\Num{5}
\end{eqnarray*}
\caption{Intermediate evaluation steps of $\hat{p}^{2}_{\Num{2}}$ }
\label{figure:intermediate-steps}
\end{figure}
gives the main intermediate steps to get to such a result.

\section{P-Time Completeness}
\label{section:P-Time Completeness}

We are now in the position to prove P-Time completeness of ILAL,
by encoding  P-Time Turing machines in $\Terms$.
We shall establish some notations together with some simplifying, but not
restricting, assumptions on the class of P-Time Turing machines we want
to encode in $\Terms$.

Every machine we are interested to is a tuple 
$\langle \mathcal{S}, \Sigma\cup\{\star,\bot,\top\}, 
\transitionf, \sfs_0, \sfs_a\rangle$, where:
\begin{itemize}
\item
$\mathcal{S}$ is the set of states with cardinality $|\mathcal{S}|$,
\item
$\Sigma$ is the \emph{input alphabet}, 
\item
$\star,\bot,\top\not\in\Sigma$ are ``\emph{blank}'' symbols,
\item
$\Sigma\cup\{\star,\bot,\top\}$ is the \emph{tape alphabet},
\item
$\transitionf$ is the \emph{transition function},
\item
$\sfs_0$ is the \emph{starting state}, and
\item
$\sfs_a$ is the \emph{accepting state}.
\end{itemize}
In general, we shall use $\sfs$ to range over $\mathcal{S}$.

The transition function has type
$\transitionf:(\Sigma\cup\{\star,\bot,\top\})\times\mathcal{S}\longrightarrow 
(\Sigma\cup\{\star,\bot,\top\})\times\mathcal{S}\times\{L,R\}$,
where $\{L,R\}$ is the set  of directions the head can move.

Both $\bot$, and $\top$ are special ``blank'' symbols. They delimit the
leftmost and the rightmost tape edge. This means that we only consider
machines with a finite tape which, however, can be extended at will.
For example, suppose the head of the machine is reading $\top$, \ie\
the rightmost limit of the tape. Assume also the head needs to move
rightward, and that, before moving, it needs to write the symbol $1$ on
the tape. Since the head is on the edge of the tape, the \emph{control} of
the machine firstly writes $1$ for $\top$, then adds a new $\top$ to the right of $1$,
and, finally, it shifts the head one place to its right, so placing the head on
the just added $\top$. The same can happen to $\bot$ when the head is
on the leftmost edge of the tape.

Obviously, the machines whose finite tape can be extended at will are
perfectly equivalent to those that, by assumption, have infinite tape.
These latter have a control that does not require to recognize the ends of the
tape, in order to extend it, when necessary.

Taking only machines with finite tape greatly simplifies our encoding, because
$\Terms$ contain only finite terms.

Recall now that we want to encode \emph{P-Time} Turing machines.
For this reason, we require that every machine comes with a polynomial
$p^\vartheta_x$, with maximal non null degree $\vartheta$. The polynomial
characterizes the maximal running time. So, every P-Time machine accepts an
input of \emph{length} $l$ if, after at most $p^\vartheta_l$ steps, it 
enters state $\sfs_a$. Otherwise, it rejects the input.

Without loss of generality, we add some further simplifying assumptions.
Firstly, whenever the machine is ready to accept the input, before entering
$\sfs_a$, it shifts its head to the leftmost tape character, different from
$\bot$. We agree that the output is the portion of tape from $\bot$, excluded,
through the first occurrence of $\star$ to its right. (Of course, for any P-Time
Turing machine there is one behaving like this with a polynomial overhead.)
Secondly, we limit ourselves to P-Time Turing machines with $\Sigma=\{0,1\}$. 

\begin{definition}
$\mathcal{T}^{\vartheta}_{\text{P-Time}}$ is the set of all P-Time Turing
machines, described here above.
\end{definition}

The next subsections introduce the parts of the encoding of a generic P-Time
Turing machine, using an instance of $\Terms$, built from the set of
variable names $\TermVars=\{0,1,\star,\bot,\top\}$. Namely, we use the symbols
of  the tape alphabet directly as \emph{variable} names for the term of the
encoding. We hope this choice will produce a clearer encoding.
We shall try to give as much intuition as possible as the development of the
encoding proceeds. However, some details will become clear
only at the end, when all the components will be assembled together.

	\subsection{States}
\label{subsection:States}

Recall that the set of states $\mathcal{S}$ has cardinality $|\mathcal{S}|$.
Assume to enumerate $\mathcal{S}$. The $i^{\text{th}}$ state is:
\begin{eqnarray*}
\State_i&=&
\lambda x_0\tensor \ldots\tensor x_{|\mathcal{S}|-1}\tensor v. x_i\ v
\qquad \text{ with } 0\leq i\leq |\mathcal{S}|-1
\enspace ,
\end{eqnarray*}
which has type:
\begin{eqnarray*}
\StateT=
\forall\A\B.
(\overbrace{(\A\linimpl\B)\tensor \ldots\tensor (\A\linimpl\B)
           }^{|\mathcal{S}| \text{ times}}
\tensor\ \A
)\linimpl\B\enspace .
\end{eqnarray*}
Every $\State_i$ extracts a row from an array that, as we shall see, encodes
the translation $\Tfun$ of $\transitionf$.
So, every $x_i$ stands for the $i^{\text{th}}$ row of $\Tfun$ which
\emph{must be} a closed term.
The parameter $v$ stands for the variables that the rows of
$\Tfun$ would \emph{share} in case they \emph{were not closed}
terms. The point here is that the sharing
is \emph{additive} and not \emph{exponential}.
We can understand the difference by assuming to apply $\State_i$ on a
$\Tfun$ with two rows $R_1$ and $R_2$.
Once all the encoding will be complete, we shall see that, as the computation
proceeds, for every instance of $\Tfun$ that the computation generates,
\emph{only one} between $R_1, R_2$ is used. 
The other gets discarded.
This has some interesting consequences on the
form of $\Tfun$ itself, if $R_1, R_2$ share some variables.
Indeed, assume
$x_1,\ldots,x_n$ be \emph{all} the free variables, with \emph{linear} types,
\emph{common} to $R_1, R_2$.
Then $R_1\otimes R_2$ can not be typed as it is: every
$x_j$ would require an \emph{exponential} type, contrasting with the
effective use of every $x_j$ we are going to do: since we assume to use
\emph{either} $R_1$, \emph{or} $R_2$, every $x_j$ is eventually
used linearly. For this reason, our instance of $\Tfun$ is represented as
the triple:
\begin{eqnarray*}
(\lambda x_1\tensor \ldots\tensor x_n.R_1)
 \otimes 
(\lambda x_1\tensor \ldots\tensor x_n.R_2)
 \otimes 
 (x_1\tensor \ldots\tensor x_n)
\enspace .
\end{eqnarray*}
The leftmost component is extracted by means of $\State_0$
that applies
$\lambda x_1\tensor \cdots\tensor x_n.R_1$ to
$x_1\tensor \cdots\tensor\; x_n$.
The rightmost component is obtained analogously, by applying
$\State_1$ to $\lambda x_1\tensor \cdots\tensor\; x_n.R_2$ to
$x_1\tensor \cdots\tensor\; x_n$. Giving linear types to
the free variables of the rows in $\Tfun$, allows their efficient,
in fact \emph{linear},  use.

	\subsection{Configurations}
\label{subsection:Configurations}

Each of them stands for the position of
the head on an instance of tape, in some state. 
We choose the following term scheme to encode the configurations of
P-Time Turing machines:
\begin{eqnarray*}
\Conf&=&
\lambda 0 1 \star \bot \top.
\\
&&
\quad
\S(\lambda x x'.
    (\DB\chi_1(\ldots(\DB\chi_{p} (\DB \bot\ x))\ldots))
    \otimes
    (\DB\chi'_1(\ldots(\DB\chi'_{q} (\DB \top\ x'))\ldots))
   \otimes 
   \State_i
  )
\enspace ,
\end{eqnarray*}
where
$\chi_{1\leq i \leq p},\chi'_{1\leq j \leq q}
 \in\{0, 1, \star\}$, with $p, q\geq 0$.
Every $\Conf$ has type:
\begin{eqnarray*}
\ConfT&=&
\forall \A.
!(\A\linimpl\A)\linimpl
!(\A\linimpl\A)\linimpl\\
&&\phantom{\forall \A.}\qquad
!(\A\linimpl\A)\linimpl
!(\A\linimpl\A)\linimpl\\
&&\phantom{\forall \A.}\qquad\qquad
!(\A\linimpl\A)\linimpl
\S(\A \linimpl \A \linimpl (\A\tensor \A\tensor \StateT))  
\enspace .
\end{eqnarray*}

As an example, take the following tape:
\begin{eqnarray}
\label{eqref:tape1}
\bot \star 1\top
\enspace .
\end{eqnarray}
Assume that the head is reading $\bot$,
and that the actual state is $\sfs_i$. Its encoding is:
\begin{eqnarray}
\label{eqref:tape1.1}
&&
\lambda 0 1 \star \bot \top.
\S(\lambda x x'.
    x
    \otimes
    (\DB \bot (\DB \star(\DB 1(\DB \top\ x'))))
    \otimes 
   \State_i
  )
\enspace .
\end{eqnarray}
The leftmost component of the tensor in the body of the
$\lambda$-abstraction is the part of the
tape to the left of the head, also called \emph{left tape}.
It is encoded in reversed order.
The cell read by the head, and the part of the tape to its right,
the \emph{right tape}, is the central component of the tensor.

Any \emph{starting configuration} has form:
\begin{eqnarray*}
\lambda 0 1 \star \bot \top.
\S(\lambda x x'. (\DB \bot\ x)
    \otimes
    (\DB \chi_1(\ldots(\DB \chi_{q} (\DB \top\ x'))\ldots))
   \otimes 
   \State_0
  )
\enspace ,
\end{eqnarray*}
where every
$\chi_j$ ranges over $\{0, 1\}$, and $\State_0$ encodes $\sfs_0$.
Namely, the tape has only characters of the input alphabet on it,
the head is on its leftmost input symbol, the left part
of the tape is empty, and the only reasonable state is the initial one.

	\subsection{Transition Function}
\label{subsection:Transition Function}

The transition function $\transitionf$ is represented by the term $\Tfun$,
which is (almost) the obvious encoding of an array in a functional language. 
So, $\Tfun$ is (essentially) a tuple of tuples. 
Every term representing a state can project a row out of $\Tfun$.
We have already seen the encoding of the states in
Subsection~\ref{subsection:States}. Since then, we know that every $\State_i$
needs as argument the set of variables additively shared by the components of
the array it is applied to. So, $\Tfun$ contains these variables as
$(|\mathcal{S}|+1)^{\text{th}}$ row.
A column of a row is extracted thanks to the projections in
Figure~\ref{figure:proj-alph-symb}.
\begin{figure}[htbp]
\begin{eqnarray*}
\Piz&=&
\lambda
0\tensor 1\tensor \star\tensor \bot\tensor \top\tensor x\tensor v.
0\ v\\
\Pio&=&
\lambda
0\tensor 1\tensor \star\tensor \bot\tensor \top\tensor x\tensor v.
1\ v\\
\Pis&=&
\lambda
0\tensor 1\tensor \star\tensor \bot\tensor \top\tensor x\tensor v.
\star\ v\\
\Pib&=&
\lambda
0\tensor 1\tensor \star\tensor \bot\tensor \top\tensor x\tensor v.
\bot\ v\\
\Pit&=&
\lambda
0\tensor 1\tensor \star\tensor \bot\tensor \top\tensor x\tensor v.
\top\ v\\
\Pie&=&
\lambda
0\tensor 1\tensor \star\tensor \bot\tensor \top\tensor x\tensor v.
x\ v
\end{eqnarray*}
\caption{Projections representing the tape alphabet symbols}
\label{figure:proj-alph-symb}
\end{figure}
The name of each projection obviously recalls the tape symbol it is associated
to. Every projection has type:
\begin{eqnarray*}
\lefteqn{\ABool_{\A,\B}=}\\
&&
((\A\linimpl\B)
 \tensor 
 (\A\linimpl\B)
 \tensor 
 (\A\linimpl\B)
 \tensor 
 (\A\linimpl\B)
 \tensor 
 (\A\linimpl\B)
 \tensor 
 (\A\linimpl\B)
 \tensor \A
)\linimpl\B
\enspace .
\end{eqnarray*}

The transition function is in Figure~\ref{figure:trans-fun}.
\begin{figure}
\begin{eqnarray*}
\Tfun&=&
\lambda
0\tensor 1\tensor\star\tensor\bot\tensor\top.
\\
&&
\begin{array}[t]{cc}
(\lambda x.Q_{0,0    }\tensor
           Q_{0,1    }\tensor
	   Q_{0,\star}\tensor
	   Q_{0,\bot }\tensor
	   Q_{0,\top }\tensor
	   Q_{0,\emptyset}\tensor
	   x)
            &\tensor \\
\vdots      &           \\
(\lambda x.Q_{|\mathcal{S}|-1,0    }\tensor
           Q_{|\mathcal{S}|-1,1    }\tensor
           Q_{|\mathcal{S}|-1,\star}\tensor
	   Q_{|\mathcal{S}|-1,\bot }\tensor
	   Q_{|\mathcal{S}|-1,\top }\tensor
	   Q_{|\mathcal{S}|-1,\emptyset}\tensor
	   x)
           &\tensor \\ 
	   & \\
0\tensor 1\tensor\star\tensor\bot\tensor\top 
\end{array}
\end{eqnarray*}
\caption{Encoding the transition function $\transitionf$}
\label{figure:trans-fun}
\end{figure}
For example, we can extract the element $Q_{i,\star}$ from $\Tfun$, 
by evaluating:
\begin{eqnarray*}
&&
\Pis\ (\State_i\ (
       \Tfun\ 
       (0     \tensor 
        1     \tensor 
        \star \tensor  
        \bot  \tensor 
        \top)
                 )
      )
\enspace .
\end{eqnarray*}

Finally, the terms $Q_{i,j}$. As expected, they produce a triple in the
codomain of the translation $\Tfun$ of $\transitionf$.
Figure~\ref{figure:output-triples}
\begin{figure}[htbp]
\begin{eqnarray*}
\Left^0_{ij}&=&
\lambda
0\tensor 1\tensor\star\tensor \bot\tensor \top .\\
&&
\qquad
\qquad
\lambda h_l t_l t_r. t_l \otimes (h_l\ (0\ t_r )) \otimes \State_{ij}
\\
\Left^1_{ij}&=&
\lambda
0\tensor 1\tensor \star\tensor \bot\tensor \top .\\
&&
\qquad
\qquad
\lambda h_l t_l t_r. t_l \otimes (h_l\ (1\ t_r )) \otimes \State_{ij}
\\
\Left^{\star}_{ij}&=&
\lambda
0\tensor 1\tensor\star\tensor \bot\tensor \top .\\
&&
\qquad
\qquad
\lambda h_l t_l t_r. t_l \otimes (h_l\ (\star\ t_r )) \otimes \State_{ij}
\\
\Left^0_{\bot ij}&=&
\lambda
0\tensor 1\tensor\star\tensor \bot\tensor \top .\\
&&
\qquad
\qquad
\lambda h_l t_l t_r. t_l \otimes (\bot\ (0\ t_r )) \otimes \State_{ij}
\\
\Left^1_{\bot ij}&=&
\lambda
0\tensor 1\tensor\star\tensor \bot\tensor \top .\\
&&
\qquad
\qquad
\lambda h_l t_l t_r. t_l \otimes (\bot\ (1\ t_r )) \otimes \State_{ij}
\\
\Left^{\star}_{\bot ij}&=&
\lambda
0\tensor 1\tensor\star\tensor \bot\tensor \top .\\
&&
\qquad
\qquad
\lambda h_l t_l t_r. t_l \otimes (\bot\ (\star\ t_r )) \otimes \State_{ij}
\\
\Right^{0}_{ij}&=&
\lambda
0\tensor 1\tensor\star\tensor \bot\tensor \top .\\
&&
\qquad
\qquad
\lambda h_l t_l t_r. 0\ (h_l\ t_l) \otimes t_r \otimes \State_{ij}
\\
\Right^{1}_{ij}&=&
\lambda
0\tensor 1\tensor\star\tensor \bot\tensor \top .\\
&&
\qquad
\qquad
\lambda h_l t_l t_r. 1\ (h_l\ t_l) \otimes t_r \otimes \State_{ij}
\\
\Right^{\star}_{ij}&=&
\lambda
0\tensor 1\tensor\star\tensor \bot\tensor \top .\\
&&
\qquad
\qquad
\lambda h_l t_l t_r. \star\ (h_l\ t_l) \otimes t_r \otimes \State_{ij}
\\
\Right^{0}_{\top ij}&=&
\lambda
0\tensor 1\tensor\star\tensor \bot\tensor \top .\\
&&
\qquad
\qquad
\lambda h_l t_l t_r. 0\ (h_l\ t_l) \otimes (\top\ t_r) \otimes \State_{ij}
\\
\Right^{1}_{\top ij}&=&
\lambda
0\tensor 1\tensor\star\tensor \bot\tensor \top .\\
&&
\qquad
\qquad
\lambda h_l t_l t_r. 1\ (h_l\ t_l) \otimes (\top\ t_r) \otimes \State_{ij}
\\
\Right^{\star}_{\top ij}&=&
\lambda
0\tensor 1\tensor\star\tensor \bot\tensor \top .\\
&&
\qquad
\qquad
\lambda h_l t_l t_r. \star\ (h_l\ t_l) \otimes (\top\ t_r) \otimes \State_{ij}
\\
\Stay^0_{ij}&=&
\lambda
0\tensor 1\tensor\star\tensor \bot\tensor \top .\\
&&
\qquad
\qquad
\lambda h_l t_l t_r. (h_l\ t_l) \otimes (0\ t_r) \otimes \State_a
\\
\Stay^1_{ij}&=&
\lambda
0\tensor 1\tensor\star\tensor\bot\tensor \top .\\
&&
\qquad
\qquad
\lambda h_l t_l t_r. (h_l\ t_l) \otimes (1\ t_r) \otimes \State_a
\\
\Stay^{\star}_{ij}&=&
\lambda
0\tensor 1\tensor\star\tensor\bot\tensor \top .\\
&&
\qquad
\qquad
\lambda h_l t_l t_r. (h_l\ t_l) \otimes (\star\ t_r) \otimes \State_a
\enspace .
\end{eqnarray*}
\caption{The output triples of $\Tfun$}
\label{figure:output-triples}
\end{figure}
defines 15 terms to encode the triples we need.
The triples are somewhat hidden in the structure of these terms.
However, such terms have
the most natural form we came up, once we choose to manipulate the
configurations of Subsection~\ref{subsection:Configurations}.

The first three ``left'' terms move the head from the top $h_l$ of the left 
tape to the top of the right tape. This move comes after the head writes
one of the symbols among $\{0,1,\star\}$ on the tape.
For example, if the written symbol is $\star$,
the new right tape becomes $h_l(\star\ t_r)$.
We recall that $\star$ (or $0$, or $1$)  replaces the symbol read before the
move. However, if the character on top of the right tape, before the move, 
was $\bot$, one of the last three ``left'' terms must be used, instead.
They put under the head the symbol which signals the end of the tape.

The ``shifting to the right'' behave almost, but not perfectly,
symmetrically.
The main motivation is that the head is assumed to read the top of the right
tape. So, when it shifts to the right only the new character that the head
writes has to be placed on the left tape. If the head was reading $\top$
before the move, another $\top$ must be added after it. This is done by the
last three ``right'' shifts.
The last three terms are used in two ways.
When the actual state of the encoded machine is $\sfs_a$ the head cannot move
anymore. This is exactly the effect of every ``stay'' term. 
For example $\Stay^1_{ij}$ must be used when we have to simulate a head reading
$1$ in the actual state $\sfs_a$: the head must rewrite $1$ without shifting.
The ``stay'' are also used as dummy terms in the ``$\emptyset$-column''
of $\Tfun$. The elements of that column will never be used because they
correspond to the move directions when the head is beyond the tape delimiters
$\bot$, and $\top$. But this can never happen.

Of course, the choice of which term in
Figure~\ref{figure:output-triples} we have to use as $Q_{i,j}$ in $\Tfun$
must be coherent with the behavior of $\transitionf$ that we want to simulate.
We shall see an explicit example about this later.

Figure~\ref{figure:typingtrans-fun}
\begin{figure}
\begin{eqnarray*}
\Ib_{\A}&=&(\A\linimpl\A)
\\
\otimes_{\A}&=&
 \Ib_{\A}\tensor 
 \Ib_{\A}\tensor 
 \Ib_{\A}\tensor 
 \Ib_{\A}\tensor 
 \Ib_{\A}
\\
\tau_\A&=&
(\Ib_\A)\linimpl
\A\linimpl
\A\linimpl
( \A\tensor \A \tensor \StateT)
\\
\RowT_\A&=&
\otimes_{\A}
\linimpl
(\ShiftT_\A \tensor 
 \ShiftT_\A \tensor 
 \ShiftT_\A \tensor 
 \ShiftT_\A \tensor 
 \ShiftT_\A \tensor 
 \ShiftT_\A
 \!\otimes
 (\otimes_{\A})
)\\
\ShiftT_\A&=&
\otimes_{\A}\linimpl\tau_\A
\\
\TfunT&=&\forall \A.
\otimes_{\A}\linimpl
(\underbrace{
\RowT_\A
  \tensor 
 \ldots
 \tensor\;
 \RowT_\A
}_{|\mathcal{S}|}
\tensor\
(\otimes_{\A}))
\\
\Left^{\chi}_{\chi'}&:& \ShiftT_\A
\\
\Right^{\chi}_{\chi'}&:& \ShiftT_\A
\\
\Stay^{\chi}_{\chi'}&:& \ShiftT_\A
\quad
\text{with } \chi\in\{0,1,\star\}\text{ and }
             \chi'\in\{\bot ij, ij\}
	     \\
\Tfun&:&\TfunT     
\enspace .
\end{eqnarray*}
\caption{Typing for $\Tfun$}
\label{figure:typingtrans-fun}
\end{figure}
gives useful hints to those who want to check the well typing of $\Tfun$.
It may help also saying that, once the whole encoding will be set up,
the projections $\Piz,\Pio,\Pis,\Pib,\Pit$, and $\Pie$
will be used in $\Tfun$ with the type instantiated as
$\ABool_{\otimes_{\A},\tau_\A}$. 

	\subsection{The Qualitative Part}
\label{subsection:Qualitative Part}

We shall use the definitions in Figure~\ref{figure:useful-defs},
\begin{figure}
\begin{eqnarray*}
\Ib_\A&=&(\A\linimpl\A)
\\
\otimes_{\A}&=&
 \Ib_{\A}\tensor 
 \Ib_{\A}\tensor 
 \Ib_{\A}\tensor 
 \Ib_{\A}\tensor 
 \Ib_{\A}
\\
!\otimes_{\A}&=&
 (!\Ib_{\A})\otimes
 (!\Ib_{\A})\otimes
 (!\Ib_{\A})\otimes
 (!\Ib_{\A})\otimes
 (!\Ib_{\A})
\\
I&=&\lambda x.x:\Ib_\A
\\
\mathsf{P}^{\otimes}&=&
 0     \,\tensor\,
 1     \,\tensor\,
 \star \,\tensor\,\,
 \bot  \,\tensor\,
 \top :\otimes_{\A}
\\
!\mathsf{P}^{\otimes}&=&
 \DB 0     \,\tensor\,
 \DB 1     \,\tensor\,
 \DB \star \,\tensor\,\,
 \DB \bot  \,\tensor\,
 \DB \top :!\otimes_{\A}
\end{eqnarray*}
\caption{Some useful definitions and abbreviations}
\label{figure:useful-defs}
\end{figure}
which also recalls some of the already introduced abbreviations.
Observe that $!\mathsf{P}^{\otimes}$ is not a term which represents a derivation of
ILAL. However, it is perfectly sensible
to associate it the logical formula that we denote by $!\otimes_{\A}$.
In particular, $!\mathsf{P}^{\otimes}$ contributes to build a well formed term, once
inserted in a suitable context.

The key terms to encode a P-Time Turing machine are in
Figure~\ref{figure:config2config}.
\begin{figure}
\begin{eqnarray*}
\Conftoconf&=&
\lambda c 0 1 \star \bot\top.
\S(
   \lambda x x'.
   (\Comp\ !\mathsf{P}^{\otimes} 
   )\\
&&\phantom{\lambda c 1 0 \star \bot\top.\S(\lambda x x'.(\ \ }
  ( \DP(c\ 
        !(\Step\ \Piz\ \DB 0)\\
&&\phantom{\lambda t 1 0 \star \bot\top.\S(\lambda x x'.(\ \ ( \DP(t\ }
        !(\Step\ \Pio\ \DB 1)\\
&&\phantom{\lambda c 1 0 \star \bot\top.\S(\lambda x x'.(\ \ ( \DP(t\ }
        !(\Step\ \Pis\ \DB \star)\\       
&&\phantom{\lambda c 1 0 \star \bot\top.\S(\lambda x x'.(\ \ ( \DP(t\ }
        !(\Step\ \Pib\ \DB \bot)\\
&&\phantom{\lambda c 1 0 \star \bot\top.\S(\lambda x x'.(\ \ ( \DP(t\ }
        !(\Step\ \Pit\ \DB \top)\\       
&&\phantom{\lambda c 1 0 \star \star'\bot\top.\S(\lambda x x'.(\ \ ( \DP }
  )(\Base\ \Pie\ x)\ (\Base\ \Pie\ x')\\
&&\phantom{\lambda c 1 0 \star \bot\top.\S(\lambda x x'.(\ \ }
   )\\
&&\phantom{\lambda c 1 0 \star \bot\top.\S }
  )
\\
\Comp&=&
\lambda \mathsf{P}^{\otimes}.
\lambda (h^l_l\tensor h^r_l\tensor  t_l)\tensor 
        (h^l_r\tensor  h^r_r\tensor t_r)\tensor s.
 h^l_r
 (s(\Tfun\ \mathsf{P}^{\otimes}))\
 h^r_l\
 t_l\
 t_r
\\
\Step&=&
\lambda xy.
 \lambda u\tensor  v\tensor  z.
   x\tensor  y\tensor  (v\ z)
\\
\Base&=&
\lambda xy. x\tensor  I\tensor  y
\end{eqnarray*}
\caption{Terms producing a configuration from another configuration}
\label{figure:config2config}
\end{figure}

$\Conftoconf$ takes a configuration $c$ and yields a new one.
Step by step, let us see the evaluation of $\Conftoconf$ applied to
the configuration \eqref{eqref:tape1.1}.
Substituting \eqref{eqref:tape1.1} for $c$,
the evaluation of the whole sub-term in the scope of the $\DP$ operator
yields:
\begin{eqnarray}
\label{eqref:int-tape}
&&
    (\Pie\tensor I \tensor x)
     \otimes
    (\Pib \tensor\DB \bot\tensor(\DB \star(\DB 1(\DB\top\ x')))
   )\otimes 
   \State_i
\enspace .
\end{eqnarray}
Observe that \eqref{eqref:int-tape} is obtained because \eqref{eqref:tape1.1}
\emph{iterates} every $\Step$  from $\Base$
in order to extract what we call \emph{head pairs} from the tape.
In this example, the two head pairs are $\Pie\tensor I$,
and $\Pib \tensor\DB \bot$. The head pairs always have the same form:
$\Piz$ will always be associated to $\DB 0$, 
$\Pio$ to $\DB 1$, 
$\Pis$ to $\DB \star$,
$\Pib$ to $\DB \bot$,
$\Pit$ to $\DB \top$, and
$\Pie$ to $I$.

Each  of $\Piz, \Pio, \ldots$, together with $\State_i$,
extracts an element in a row of $\Tfun$.
This happens in $\Comp$. In its body, the actual state $\sfs$ extracts
a row from $\Tfun$, and the tape symbol $h^l_r$, read by the head,
picks a move out of the row.
In our running  example, $s$ is $\State_i$, and $h^l_r$ is $\Pib$.
So, if $\transitionf(\sfs_i,\bot)=(\sfs_j,1,L)$, then 
$Q_{i,\bot}$, producing $(\sfs_j,1,L)$, must be $\Left^{1}_{\bot ij}$.
The next computational steps are, internal to $\Comp$ are:
\begin{eqnarray*}
\lefteqn{
\Pib\ (\State_i\ (\Tfun\ !\mathsf{P}^{\otimes}))\ I\ 
					   x\ 
					   (\DB \star(\DB 1(\DB \top\ x')))
	}
\\
&\qquad\qquad\rewsyst^*&
\Left^{1}_{\bot ij}\ !\mathsf{P}^{\otimes}\ I\ 
			             x\ 
				     (\DB \star(\DB 1(\DB \top\ x')))
\\
&\qquad\qquad\rewsyst^*&
(\lambda h_l t_l t_r. t_l \otimes
                      (\DB \bot(\DB 1\ t_r))\otimes
		      \State_{j})\ I\ 
			             x\ 
				     (\DB \star(\DB 1(\DB \top\ x')))
\\
&\qquad\qquad\rewsyst^*&
x \otimes 
(\DB \bot\ (\DB 1(\DB \star(\DB 1(\DB \top\ x'))))) \otimes
\State_{j}
\end{eqnarray*}

So, under the hypothesis of simulating
$\transitionf(\sfs_i,\bot)=(\sfs_j,1,L)$, the term $\Conf$ rewrites
\begin{eqnarray}
\label{eqref:tape1.2}
\lambda 0 1 \star \bot \top.
\S(\lambda x x'.
    x \otimes
    (\DB \bot (\DB \star(\DB 1(\DB \top\ x'))))
    \otimes 
   \State_i
  )
\end{eqnarray}
into:
\begin{eqnarray}
\label{eqref:tape1.3}
\lambda 0 1 \star \bot \top.
\S(\lambda x x'.
    x \otimes
    (\DB \bot (\DB 1 (\DB \star(\DB 1(\DB \top\ x')))))
    \otimes 
   \State_j
  )
\end{eqnarray}
by means of $\Conftoconf$.
For those who want to check that $\Conftoconf$ is \emph{iterable},
\ie\ that $\Conftoconf:\ConfT\linimpl\ConfT$,
Figure~\ref{figure:iterableconf} gives some useful hints on the typing.
\begin{figure}
\begin{eqnarray*}
\BaseT
&=&
\forall \A \B. \ABool_{\A,\B}\linimpl
\A\linimpl\ABool_{\A,\B}
\\
\StepT
&=&
\forall \A \B. \ABool_{\A,\B}\linimpl
\Ib_\A\linimpl
\ABool_{\A,\B}\linimpl
\ABool_{\A,\B}
\\
\CompT
&=&
\forall \A .
\otimes_{\A}\linimpl
((\ABool_{\otimes_{\A},\tau_{\A}}\tensor
 (\Ib_{\A})\tensor
 \A)\tensor
\\
&&\phantom{\forall \A . \otimes_{\A}\linimpl(}
 (\ABool_{\otimes_{\A},\tau_{\A}}\tensor
 (\Ib_{\A})\tensor
 \A)
\tensor
\\
&&\phantom{\forall \A .\otimes_{\A}\linimpl (}
\StateT
)
\linimpl
(\A\tensor \A\tensor \StateT)
\\
\Base&:&\BaseT
\\
\Step&:&\StepT
\\
\Comp&:&\CompT
\end{eqnarray*}
\caption{Typing for $\Conftoconf$}
\label{figure:iterableconf}
\end{figure}

	\subsection{The Whole Encoding}
\label{subsection:Whole Encoding}

We are, finally, in the position to complete our encoding of the
machines in $\mathcal{T}^{\vartheta}_{\text{P-Time}}$, with a given
$\vartheta$, as derivations of ILAL.

Up to now, we have built the two  main parts of the encoding.
We call them \emph{qualitative}, and \emph{quantitative}.
The encoding $\Tfun$ of the transition function, and the iterable term
$\Conftoconf$, which maps configurations to configurations, belong to
the first part. The encoding of the polynomials falls into the latter.

The whole encoding exploits the quantitative part to iterate
the qualitative one, starting from the
initial configuration. This is a suitable extension
of the actual input. Every actual input of the encoding is a list, standing for
a tape with the symbols $\{0,1\}$ on it. The iteration is as long as the value
of the encoding of the polynomial, applied to the (unary representation)
of the length of the actual input.

\begin{theorem}
There is a translation 
$\Hat{\phantom{T}}:\mathcal{T}^{\vartheta}_{\text{P-Time}}\rightarrow\Terms$
such that, for any $T\in\mathcal{T}^{\vartheta}_{\text{P-Time}}$, and any input
stream $x$ for $T$, if $T x$ evaluates to $y$, then
$\Hat{T}\Hat{x}\rewsyst^*\Hat{y}$. In particular,
$\Hat{T}:\TapeT\linimpl\S^{\vartheta+6}\TapeT$,
where:
\begin{eqnarray*}
\TapeT&=&
\forall \A.
!(\A\linimpl\A)
\linimpl
!(\A\linimpl\A)
\linimpl
\S(\A\linimpl\A)
\enspace .
\end{eqnarray*}
\end{theorem}

The rest of this subsection develops the details about 
$\Hat{\phantom{T}}:\mathcal{T}^{\vartheta}_{\text{P-Time}}\rightarrow\Terms$.

Figure~\ref{figure:input-encoding} 
\begin{figure}[htbp]
\begin{eqnarray*}
&&
\lambda 01.
\S(\lambda x.
   \DB\chi_1(\ldots(\DB\chi_{p}\ x)\ldots)
  )
\enspace ,
\\
&&\text{where }
\chi_{1\leq i \leq p},\in\{0, 1\} \text{ and } p\geq 0
\enspace .
\end{eqnarray*}
\caption{Encoding the input tapes}
\label{figure:input-encoding}
\end{figure}
introduces the general scheme to encode any
input for $T$ as a term.

Figure~\ref{figure:root-whole-encoding}
\begin{figure}[htbp]
\begin{eqnarray*}
\Hat{T}
&=&
\lambda t.
\Conftotape^{\vartheta+5} 
(\S(
    (\lambda t_1\tensor t_2.
      \Iterint^{\vartheta+3}\
      (\Hat{p}^{\vartheta}_{x}(\Tapetoint\ t_1))\\
&&\phantom{\lambda t. \Conftotape^{\vartheta+3} (\S( (\lambda t_1\tensor t_2.
           \Iterint^{\vartheta+3}\
          }
      (! \Conftoconf)\\
&&\phantom{\lambda t. \Conftotape^{\vartheta+3} (\S( (\lambda t_1\tensor t_2.
           \Iterint^{\vartheta+3}\
          }
      (\Tapetoconf\ t_2)\\
&&\phantom{\lambda t. \Conftotape^{\vartheta+3} (\S( }
  )\ \DP(\Doubletape\ t))): \TapeT\linimpl\S^{\vartheta+5}\TapeT
\end{eqnarray*}
\caption{Encoding a P-Time Turing machine in
$\mathcal{T}^{\vartheta}_{\text{P-Time}}$}
\label{figure:root-whole-encoding}
\end{figure}
shows the encoding $\Hat{T}$ of $T\in\mathcal{T}^{\vartheta}_{\text{P-Time}}$
which glues the quantitative and the qualitative parts together.

Figures~\ref{figure:doubles-tape},
\begin{figure}[htbp]
\begin{eqnarray*}
\Doubletape&=&
\lambda t.
\S( \DP(t\ !(\lambda x y.(\Succtape_0\ x)\tensor(\Succtape_0\ y)
            )\\
&&\phantom{\lambda t.\S( \DP(t\ }
           !(\lambda x y.(\Succtape_1\ x)\tensor(\Succtape_1\ y)
	    ) \\
&&\phantom{\lambda t.\S( \DP }
        )\ \Emptytape\ \Emptytape
   ):\TapeT\linimpl\S(\TapeT\tensor\TapeT)\\
\text{where}&&\\
\Succtape_{\chi}&=&
\lambda t01.
\S(\lambda x.\DB\chi\ \DP(t\ 0\ 1)\ x):
\TapeT\linimpl\TapeT\qquad\text{ with }\chi\in\{0,1\}
\\
\Emptytape&=&
\lambda 01.\S(\lambda x.x):\TapeT
\end{eqnarray*}
\caption{Doubling the contents of the actual input tape}
\label{figure:doubles-tape}
\end{figure}
\ref{figure:configuration-to-tape}, 
\begin{figure}[htbp]
\begin{eqnarray*}
\Conftotape^p&=&
\lambda c.\S^p((\lambda 01.\S((\lambda w\tensor y\tensor z.y)\\
&&
\phantom{\lambda c.\S^p((\lambda 01.\S( }
					(\DP(\DP^p c\ 0\ 1\ 
					              !(\lambda w. \Emptytape)\ !I\
						      !I)\\
&&
\phantom{\lambda c.\S^p((\lambda 01.\S( (\DP(}
				         \Emptytape\ 
					 \Emptytape))\\
&&
\phantom{\lambda c.\S^p((}
			     )\ !\Succtape_0\ !\Succtape_1\\
&&
\phantom{\lambda c.\S^p (}		     
	      ):\S^{p}\ConfT\linimpl\S^{p+1}\TapeT
\end{eqnarray*}
\caption{Reading back a tape from a configuration}
\label{figure:configuration-to-tape}
\end{figure}
\ref{figure:tape-to-initial-configuration},
\begin{figure}[htbp]
\begin{eqnarray*}
\Tapetoconf&=&
\lambda t.\Coercinitconf\ (\Tapetoinitconf\ t):
\TapeT\linimpl\S\ConfT
\\
\text{where:}&&
\\
\Coercinitconf&=&
\lambda c.
\S( \DP(c\ !(\Succinitconf_0)\\
&&\phantom{\lambda c.\S( \DP(c\ }
           !(\Succinitconf_1)\\
&&\phantom{\lambda c.\S( \DP(c\ }
           !(\Succinitconf_{\top})\\
&&\phantom{\lambda c.\S( \DP}
        )\ \Emptyinitconf\ \Emptyinitconf\\
&&\phantom{\lambda c.\S(}
  ):\ConfT\linimpl\S\ConfT
\\
\Emptyinitconf&=&
\lambda 01\star\bot\top. \S(\lambda x x '. (\DP\bot\ x)\tensor(\DP\top\ x'))
:\ConfT\\
\Succinitconf_{\chi}&=&
\lambda c 01\star\bot\top.
\S(\lambda x x '. (\lambda w\tensor w'\tensor s.
                   w \tensor (\DB\chi\ w')\tensor s)
\\
&&\phantom{\lambda c01\star\bot\top.\S(\lambda x x '. 
                                      (\lambda w\tensor w'\tensor s. }
                  (\DP(c\ 0\ 1\ \star\ \bot\ \top)\ x\ x')\\
&&\phantom{\lambda c01\star\bot\top.\S }
  ):\ConfT\linimpl\ConfT\qquad\text{ where } \chi\in\{0,1\}
\\
\text{and:}&&
\\
\Tapetoinitconf&=&
\lambda t01\star\bot\top.
\S(\lambda w w'. \Step\ (\DB\bot\ w)\\
&&\phantom{\lambda t01\star\bot\top.\S(\lambda w w'. \Step\ }
                        \DB(t\ !(\Step \DB\ 0)\\
&&\phantom{\lambda t01\star\bot\top.\S(\lambda w w'. \Step\ \DB(t\ }
	 	               !(\Step \DB\ 1)\\
&&\phantom{\lambda t01\star\bot\top.\S(\lambda w w'. \Step\ \DB }
		           )\ I\tensor (\DB\top\ w')\tensor \State_0\\
&&\phantom{\lambda t01\star\bot\top.\S( }
  ):\TapeT\linimpl\ConfT
\\
\Step&=&
\lambda x. \lambda y\tensor z. x \tensor(y\ w):
\B\linimpl((\A\linimpl\A)\tensor\A)\linimpl(\B\tensor\A)
\end{eqnarray*}
\caption{The initial configuration out of the actual input tape}
\label{figure:tape-to-initial-configuration}
\end{figure}
\ref{figure:tape-to-integer},
\begin{figure}[htbp]
\begin{eqnarray*}
\Tapetoint&=&
\lambda ts.\S(\lambda x.\DP(t\ s\ s)\ x)
:\TapeT\linimpl\IntT
\end{eqnarray*}
\caption{Transforming the actual input tape into an integer}
\label{figure:tape-to-integer}
\end{figure}
and \ref{figure:general-iteration}
\begin{figure}[htbp]
\begin{eqnarray*}
\Iter^p&=&
\lambda xyz.
\S^p(\S(\DP(\DP^p x\ y)\ \DP z
           )
    )
:\IntT^p\linimpl !(A\linimpl A)\linimpl\S A\linimpl\S^{p+1}A
\end{eqnarray*}
\caption{Generalizing the iteration}
\label{figure:general-iteration}
\end{figure}
introduce the terms $\Doubletape$, $\Conftotape$, $\Tapetoinitconf$,
$\Tapetoint$, and the generalization $\Iter^p$ of $\Iter$, with $1\leq p$,
used by $\Hat{T}$.

The term $\Doubletape$, applied to a tape, doubles it. This is possible only by
accepting that the result gets embedded into a $\S$-box.
For example:
\begin{eqnarray*}
\lefteqn{
\Doubletape\
(\lambda 01.\S(\lambda x. \DB 1(\DB 0\ x)
	      )
)
}\\
&\rewsyst^*&
\S
(
(\lambda 01.\S(\lambda x. \DB 1(\DB 0\ x)
	      )
)
\tensor
(\lambda 01.\S(\lambda x. \DB 1(\DB 0\ x)
	      )
)
)
\enspace .
\end{eqnarray*}

The term $\Conftotape$ is used to erase the garbage, left by $\Hat{T}$ on its
tape, to produce the result. Recall, indeed, that we made some assumptions on
the behavior of the elements of $\mathcal{T}^{\vartheta}_{\text{P-Time}}$
when entering $\sfs_a$. The hypothesis was that the
machines we encode enter $\sfs_a$ after their heads read the
leftmost element of the tape, different from $\bot$.
A further assumption is that the result
is the portion of tape falling between the head position and the first
occurrence of $\star$ to its right, once the machine is in state $\sfs_a$, 
The term $\Conftotape$ eliminates all the components of
the encoding of a tape which is $p$ $\S$-boxes deep, but those between $\bot$,
and the leftmost occurrence of $\star$. For example, if $\Hat{T}$ reaches the
configuration:
\begin{eqnarray*}
C&=&
\S^p(\lambda 01\star\bot\top.
\S(\lambda x x'. \DB\bot\ x \tensor
                 \DB 1(\DB \star(\DB 0(\DB\top x')))\tensor
		 \State_a))
		 \enspace,
\end{eqnarray*}
then $\Conftotape^p\ C\rewsyst^* \S^{p+1}(\lambda 01.\S(\lambda x.\DP 1\ x))$,
\ie\ the result of the simulated machine is simply the tape with the single
alphabet element $1$, and embedded into $p+1$ $\S$-boxes.

The term $\Tapetoconf$ goes in the opposite direction than
$\Conftotape$. Given the encoding of a tape $t$, $\Tapetoconf\ t$
gives the initial configuration of the encoded machine, embedded into one
$\S$-box. For example:
\begin{eqnarray*}
\lefteqn{
\Tapetoconf\
(\lambda 01.\S(\lambda x. \DB 1(\DB 0\ x)
	      )
)
}\\
&\rewsyst^*&
\S(\lambda 01\star\bot\top.
  \S(\lambda x x'. 
     \DB\bot\ x
     \tensor
     \DB 1(\DB 0(\DB \top\ x))
     \tensor
     \State_0
    )
  )
\enspace .
\end{eqnarray*}

The term $\Tapetoint$, applied to a tape, produces the numeral,
which expresses the unary length of the tape itself. For example:
\begin{eqnarray*}
\Tapetoint\
(\lambda 01.\S(\lambda x. \DB 1(\DB 0\ x)
	      )
)
&\rewsyst^*&
\lambda y.\S(\lambda x. \DB y(\DB y\ x))
\enspace .
\end{eqnarray*}

The term $\Iter^p$ is the obvious generalization of $\Iter$ to a first
argument with type $\IntT^p$.

As a summary, we rephrase the intuitive explanation we gave at the beginning
of this subsection, to describe the behavior of the encoding.
$\Iter^{\vartheta+3}$ iterates $\Hat{p}^{\vartheta}_x\ (\Tapetoint\ t_1)$ times
the term $!\Conftoconf$, starting from the initial configuration
given by $\Tapetoconf\ t_2$.
The variables $t_1, t_2$ stand for the two copies of the input tape, produced by
$\Doubletape\ t$, where $t$ represents the input tape itself.
Finally, $\Conftotape^{\vartheta+6}$ reads back the result.

\section{Conclusions}
\label{section:Conclusions}

Light Linear Logic \cite{Girard:LLL98} is the first logical system with cut
elimination, whose formulas can be used as program annotations
to improve the evaluation efficiency of the reduction. 
In the remark concluding
Subsection~\ref{subsection:The Predecessor}, 
we observed that the relation between the strategy to get such an 
efficiency and the more traditional strategies is not completely clear;
we left an open problem.

By drastically simplifying Light Linear Logic sequent calculus,
Light Affine Logic helps to understand the main crucial
issues of Girard's technique to control the computational complexity. 
Roughly, it can be summarized in the motto:
\emph{stress and take advantage of linearity whenever possible}.
Technically, the simplification allows to see $\S$ as a \emph{weak} version of 
dereliction in Linear Logic \cite{Gi95}. It opens $!$-boxes while preserving the
information on levels.
Moreover, P-Time completeness has not a completely trivial proof.
In particular, some reader may have noticed that the
configurations of the machines are not encoded obviously, like in 
\cite{Girard:LLL98}, as recalled in Figure~\ref{figure:Obvious Encoding}.
\begin{figure}[htbp]
\begin{eqnarray*}
&
\lambda 0 1 \star \bot.
\S(\lambda x.
    (\DB\chi_1(\ldots(\DB\chi_{p} (\DB \bot\ x))\ldots))
  )
&
\\
&
\otimes
&
\\
&
\lambda 0 1 \star \top.
\S(\lambda y.
    (\DB\chi'_1(\ldots(\DB\chi'_{q} (\DB \top\ y))\ldots))
  )
&
\\
&
\otimes
&
\\
& 
\State_i
\end{eqnarray*}
where
$\chi_{1\leq i \leq p},\chi'_{1\leq j \leq q}
 \in\{0, 1, \star\}$, with $p, q\geq 0$.
\caption{Obvious encoding of the configurations}
\label{figure:Obvious Encoding}
\end{figure}
\cite{Roversi:CSL99} discusses about why such an encoding can not work.
Roughly, it does not allow to write an \emph{iterable} function $\Conftoconf$,
which is basic to produce the whole encoding.

The idea to consider full weakening in Light Linear Logic, to get Light Affine
Logic, was suggested by the fact that
in Optimal Reduction \cite{AG98} we may freely erase any term.
For the experts: the garbage nodes do not get any index. 

Some attempts to extract a programming language
with automatic polymorphic type inference,
from ILAL are in \cite{Roversi:ASIAN98,Roversi:IJFCS00}.
However, they must be improved in terms of expressivity and readability.

Finally, it would be interesting tracing some relation between Light Affine
Logic and other languages that characterize P-Time, like, just to make an
example Bellantoni-Cook system in \cite{BellantoniCook:CC92}.


\begin{thebibliography}{99}
\bibitem[Asp98]{Asperti:LICS98}
A.~Asperti.
\newblock Light {A}ffine {L}ogic.
\newblock In {\em Proceedings of Symposium on Logic in Computer Science
  LICS'98}, 1998.

\bibitem[AM98]{AM98} A. Asperti, H.Mairson {\em Optimal
$\beta$-reduction is not elementary recursive}. 
Proc. of the twenty-fifth  Annual ACM SIGACT-SIGPLAN Symposium 
on Principles of Programming Languages (POPL'98).

\bibitem[AG98]{AG98} A. Asperti, S.Guerrini {\em The
Optimal Implementation of Functional Programming Languages}.
To appear in the {\em ``Cambridge Tracts in Theoretical Computer
Science''} Series, Cambridge University Press, 1998.

\bibitem[BC92]{BellantoniCook:CC92}
S.~Bellantoni and S.~Cook.
\newblock A new recursion-theoretic characterization of the polytime functions.
\newblock {\em Computational Complexity}, 2:97 -- 110, 1992.

\bibitem[DJ99]{DanosJoinet:ICC99}
V.~Danos, V. and J.-B.~Joinet.
{\em Linear Logic \& Elementary Time},
First international workshop on Implicit Computational 
Complexity-1999 (ICC'99), 1999.

\bibitem[Gi95]{Gi95} 
J.-Y.~Girard. 
{\em Proof Nets: the parallel syntax for proof-theory}, 
in Ursini and Agliano, editors,  {\em Logic and Algebra}, 
Marcel Dekker, New York, 1995.

\bibitem[Gi98]{Girard:LLL98}
J.-Y.~Girard.
\newblock Light {L}inear {L}ogic.
\newblock {\em Information and Computation}, 143:175 -- 204, 1998.

\bibitem[GLT89]{GLT:PT}
J.-Y.~Girard, Y.~Lafont, and P.~Taylor.
\newblock {\em Proofs and Types}.
\newblock Cambridge University Press, 1989.

\bibitem[GSS92]{GSS92} 
J.-Y.Girard, A.Scedrov, and P.J.Scott. 
{\em Bounded Linear Logic: a modular approach to polynomial time computability},
 Theoretical Computer Science, 97:1-66, 1992.

\bibitem[KOS97]{KOS97} 
M.L.Kanovich, M.Okada, and A.Scedrov. 
{\em Phase Semantics for Light Linear Logic}.
To appear in Theoretical Computer Science, V.405. 1997.

\bibitem[Le94]{Le94} D.~Leivant.
{\em A foundational delineation of poly-time},
Information and Computation, 110:391-420, 1994.

\bibitem[LM93]{LM93} D.~Leivant, and J-Y.~Marion.
{\em Lambda Calculus characterizations of poly-time}, 
Fundamenta Informaticae, 19:167-184, 1993.

\bibitem[Mit88]{Mitchell:InfComp88}
J.C.~Mitchell.
\newblock Polymorphic type inference and containment.
\newblock {\em Information and Computation}, 76:211 -- 249, 1988.

\bibitem[Rov98]{Roversi:ASIAN98}
L.~Roversi.
\newblock A polymorphic language which is typable and poly-step.
\newblock In {\em Advances in Computing Science -- ASIAN'98}, volume LNCS 1538,
  pages 43 -- 60. Springer-Verlag, 8 -- 10 December (Manila -- The Philippines)
  1998.

\bibitem[Rov99]{Roversi:CSL99}
L.~Roversi.
\newblock A P-Time completeness proof for light logics.
\newblock In {\em Proceedings of Computer Science Logic 1999 (CSL'99) (Madrid
  -- Spain)}, volume LNCS 1683,pages 469 -- 483.
Springer-Verlag, 20 -- 25 September (Madrid -- Spain) 1999.

\bibitem[Rov00]{Roversi:IJFCS00}
L.~Roversi.
\newblock Light Affine Logic as a Programming Language:
           a First Contribution.
\newblock {\em International Journal of Foundations of Computer Science},
11(1):113 -- 152, 2000.

\end{thebibliography}
\end{document}